\documentclass{LMCS}

\def\doi{9(1:15)2013}
\lmcsheading%
{\doi}
{1--23}
{}
{}
{Sep.~28, 2012}
{Mar.~28, 2013}
{}

\usepackage{enumerate,hyperref}

\usepackage{amsmath,amscd}
\usepackage{amssymb}
\usepackage{ifthen}

\usepackage{comment}

\DeclareSymbolFont{operators}{OT1}{ptm}{m}{n}
\DeclareSymbolFont{Greek}{OT1}{cmr}{m}{n}
\DeclareMathSymbol{\Gamma}{\mathord}{Greek}{"00}
\DeclareMathSymbol{\Delta}{\mathord}{Greek}{"01}
\DeclareMathSymbol{\Theta}{\mathord}{Greek}{"02}
\DeclareMathSymbol{\Lambda}{\mathord}{Greek}{"03}
\DeclareMathSymbol{\Xi}{\mathord}{Greek}{"04}
\DeclareMathSymbol{\Pi}{\mathord}{Greek}{"05}
\DeclareMathSymbol{\Sigma}{\mathord}{Greek}{"06}
\DeclareMathSymbol{\Upsilon}{\mathord}{Greek}{"07}
\DeclareMathSymbol{\Phi}{\mathord}{Greek}{"08}
\DeclareMathSymbol{\Psi}{\mathord}{Greek}{"09}
\DeclareMathSymbol{\Omega}{\mathord}{Greek}{"0A}

\renewcommand{\nc}{\newcommand}
\nc{\rnc}{\renewcommand} \nc{\nev}{\newenvironment}
\rnc{\subsection}{\secdef\ssa\ssb}
\nc{\ssa}[2][default]{\par\vspace{1ex}\refstepcounter{subsection}\noindent\textbf{\thesubsection.
#1. }} \nc{\ssb}[1]{\par\vspace{2ex}\noindent\textbf{#1. }}

\rnc{\subsubsection}{\secdef\sssa\sssb}
\nc{\sssa}[2][default]{\par\vspace{1ex}\refstepcounter{subsubsection}\noindent\textit{\thesubsubsection.
#1. }} \nc{\sssb}[1]{\par\vspace{1ex}\noindent\textit{#1. }}

\makeatletter \rnc{\@seccntformat}[1]{{\normalfont\bfseries{\csname
the#1\endcsname}\hspace{1pt}.\hspace{0.4em}}}
\rnc{\section}{\@startsection
        {section}%
        {1}%
        {0mm}%
        {-\baselineskip}%
        {0.5\baselineskip}%
        {\normalfont\normalsize\bfseries\centering}%
}
\renewcommand{\@makecaption}[2]{\begin{center}#1. #2\end{center}}
\makeatother

\newcounter{theo}
\rnc{\thetheo}{\arabic{theo}}

\nev{defform}[2]{\refstepcounter{theo}\begin{list}{}{%
      \setlength{\labelwidth}{0em}%
      \setlength{\leftmargin}{0em}%
      \setlength{\itemindent}{\labelsep}%
      \setlength{\listparindent}{1.5em}
      \setlength{\parsep}{0em}}%
      \item[\textbf{#2 \thetheo\ifthenelse{\equal{#1}{}}{}{ (#1)}.}]}%
    {\end{list}}

\nev{satzform}[2]{\begin{defform}{#1}{#2}\itshape}{\end{defform}}

\nev{theo}[1][]{\begin{satzform}{#1}{Theorem}}{\end{satzform}}
\nev{prov}[1][]{\begin{satzform}{#1}{Proviso}{}}{\end{satzform}}
\nev{defn}[1][]{\begin{defform}{#1}{Definition}}{\end{defform}}

\nc{\proofend}{\qed}

\rnc{\labelenumi}{(\arabic{enumi})} \rnc{\labelitemi}{\text{--}}
\rnc{\phi}{\varphi} \rnc{\epsilon}{\varepsilon}
\nc{\bigmid}{\;\big|\;} \nc{\Bigmid}{\;\Big|\;}
\rnc{\max}{\textup{max}} \rnc{\min}{\textup{min}}
\rnc{\log}{\textup{log}\;}

\newlength{\probwidth}
\rnc{\prob}[3][9]{
\begin{center}
  \normalfont\fbox{
   \begin{tabular}[t]{
     rp{#1cm}}\textit{Instance:}&#2. \\
     \textit{Problem:}&#3
   \end{tabular}}
\end{center}}

\nc{\pprob}[4][9]{
\begin{center}
   \normalfont\fbox{
    \begin{tabular}[t]{
     rp{#1cm}}\textit{Instance:}&#2. \\
     \textit{Parameter:}&#3. \\
     \textit{Problem:}&#4
   \end{tabular}}
\end{center}}

\nc{\nprob}[4][9]{
\begin{center}
  \normalfont\fbox{

\addtolength{\probwidth}{#1cm}\parbox{\probwidth}{\textsc{#2}\\\hspace*{1.5em}
     \begin{tabular}[t]{
      rp{#1cm}}\textit{Instance:}&#3. \\
      \textit{Problem:}&#4
     \end{tabular}}}
\end{center}}

\nc{\npprob}[5][9]{
\begin{center}
  \normalfont\fbox{

\addtolength{\probwidth}{#1cm}\parbox{\probwidth}{\textsc{#2}\\\hspace*{1.5em}
    \begin{tabular}[t]{
     rp{#1cm}}\textit{Instance:}&#3. \\
     \textit{Parameter:}&#4. \\
     \textit{Problem:}&#5
    \end{tabular}}}
\end{center}}

\nc{\nppxrob}[5][9]{ \normalfont\fbox{

\addtolength{\probwidth}{#1cm}\parbox{\probwidth}{\textsc{#2}\\\hspace*{1.5em}
   \begin{tabular}[t]{
    rp{#1cm}}\textit{Instance:}&#3. \\
    \textit{Parameter:}&#4. \\
    \textit{Problem:}&#5
   \end{tabular}}}}

\nc{\nppprob}[5][4]{
\begin{center}
  \normalfont\fbox{

\addtolength{\probwidth}{#1cm}\parbox{\probwidth}{\textsc{#2}\\\hspace*{1.5em}
    \begin{tabular}[t]{
     rp{#1cm}}\textit{Instance:}&#3. \\
     \textit{Parameter:}&#4. \\
     \textit{Problem:}&#5
    \end{tabular}}}
\end{center}}

\nc{\FOR}{\textbf{for}} \nc{\FORALL}{\textbf{for all}}
\nc{\TO}{\textbf{to}} \nc{\DO}{\textbf{do}} \nc{\OD}{\textbf{od}}
\nc{\IF}{\textbf{if}} \nc{\FI}{\textbf{fi}}
\nc{\THEN}{\textbf{then}} \nc{\ELSE}{\textbf{else}}
\nc{\WHILE}{\textbf{while}} \nc{\REPEAT}{\textbf{repeat}}
\nc{\UNTIL}{\textbf{until}} \nc{\OR}{\textbf{or}}
\nc{\AND}{\textbf{and}} \nc{\PRINT}{\textbf{print}}

\nc{\im}[1]{\item\hspace{#1cm}}
\nev{algorithm}{\begin{enumerate}\rnc{\labelenumi}{\textit{\small
\arabic{enumi}.}}\rnc{\itemsep}{0ex}}{\end{enumerate}}

\nc{\A}{\mathfrak{A}}
\nc{\B}{\mathfrak{B}}
\nc{\C}{\mathfrak{C}}

\newcommand{\Mod}{ \ \textup{mod}\  }

\nc{\rand}[1]{\marginpar{\raggedright\footnotesize #1}}
\nc{\hrand}[1]{\rand{\textbf{H: }#1}}
\nc{\mrand}[1]{\rand{\textbf{M: }#1}}
\newcommand{\csp}{\mathsf{CSP}}
\newcommand{\qcsp}{\mathsf{QCSP}}
\newcommand{\expos}{\mathsf{EXPOS}}
\newcommand{\efpos}{\mathsf{EFPOS}}

\begin{document}

\title[An Algebraic Preservation Theorem]{An Algebraic Preservation
  Theorem 
for \texorpdfstring{$\aleph_0$}{aleph0}-Categorical
Quantified Constraint Satisfaction}


\author[H.~Chen]{Hubie Chen\rsuper a}	
\address{{\lsuper a}%
Departamento LSI, Facultad de Inform\'{a}tica,
Universidad del Pa\'{i}s Vasco,
E-20018 San Sebasti\'{a}n,
Spain;
and
IKERBASQUE, Basque Foundation for Science,
E-48011 Bilbao,
Spain
}	
\email{hubie.chen@ehu.es}

\thanks{{\lsuper a}%
The first author
was partially supported by the Spanish program ``Ramon y Cajal'' and
MICINN grant TIN2010-20967-C04-02.
The first author was also supported by
by Spanish Project FORMALISM (TIN2007-66523), 
by the Basque Government Project S-PE12UN050(SAI12/219), and 
by the University of the Basque Country under grant UFI11/45.
}

\author[M.~M\"{u}ller]{Moritz M\"{u}ller\rsuper b}	
\address{{\lsuper b}%
Kurt G\"{o}del Research Center, Universit\"{a}t Wien, Austria}
\email{moritz.mueller@univie.ac.at}

\thanks{%
{\lsuper b}The second author thanks the FWF (Austrian Science
Fund) for its support through Project P 24654 N25.
}

\ACMCCS{[{\bf  Theory of computation}]:  Logic---Constraint and logic programming} 
\titlecomment{{\lsuper*}
An extended abstract of this work appears
in the proceedings of the 2012 ACM/IEEE Symposium on
Logic in Computer Science~\cite{ChenMueller12-preservation}.}

\begin{abstract}
We prove an algebraic preservation theorem for positive Horn definability
in $\aleph_0$-categorical structures. In particular, we define and study
a construction which we call the \emph{periodic power}
of a structure, and define a \emph{periomorphism} of a structure
to be a homomorphism from the periodic power of the structure to
the structure itself.  Our preservation theorem states that,
over an $\aleph_0$-categorical structure, a relation is
positive Horn definable if and only if it is preserved by all periomorphisms
of the structure.  We give applications of this theorem, including
a new proof of the known complexity classification of
quantified constraint satisfaction  on equality templates.
\end{abstract}

\maketitle

\section{Introduction}

Model checking -- deciding if a logical sentence holds on a
structure -- is a basic computational problem which is in general
intractable; for example, model checking first-order sentences on
finite structures is well-known to be PSPACE-complete. In the
context of model checking, fragments of first-order logic based on
restricting the connectives
$\{ \wedge, \vee, \neg \}$
and quantifiers
$\{ \exists, \forall \}$
 have been
considered in a variety of settings. For instance, the problem of
model checking \emph{primitive positive} sentences, sentences formed
using $\{ \wedge, \exists \}$, is a NP-complete problem that is a
formulation of the \emph{constraint satisfaction problem (CSP)}, and
admits a number of other natural characterizations, as shown in
the classical work of Chandra and Merlin~\cite{ChandraMerlin77-optimal}.
The problem of
model checking \emph{positive Horn} sentences, sentences formed
using $\{ \wedge, \exists, \forall \}$, is known as the
\emph{quantified constraint satisfaction problem (QCSP)}, and is
PSPACE-complete; indeed, certain cases of this problem are canonical
complete problems for
PSPACE~\cite[Chapter 19]{Papadimitriou95-complexity}.
Another natural fragment consists of
 the \emph{existential positive} sentences, which are formed from
$\{ \wedge, \vee, \exists \}$.

Such syntactically restricted fragments of first-order logic can
be naturally parameterized by the
 structure~\cite{Martin08-FOparameterizedbymodel}.
As examples, consider the following problems for a structure $\A$:
\begin{itemize}

\item $\csp(\A)$: decide the primitive positive theory of $\A$.
\item $\qcsp(\A)$: decide the positive Horn theory of $\A$.
\item $\expos(\A)$: decide the existential positive theory of $\A$.
\item $\efpos(\A)$: decide the equality-free positive theory of $\A$.

\end{itemize}
Via this parameterization, one obtains four \emph{families} of problems,
and is prompted with classification programs: for each
of the families, classify the problems therein according to
their computational complexity.
On finite structures, comprehensive classifications are known
for the families $\expos(\A)$ and $\efpos(\A)$.
Each
problem $\expos(\A)$ is either in L or
NP-complete~\cite{BodirskyHermannRichoux09-existentialpositive},
and each problem $\efpos(\A)$ is either in L,
NP-complete, coNP-complete, or
PSPACE-complete~\cite{MadelaineMartin11-tetrachotomyPFOwithouteq}.
Moreover, each of these two classifications is effective
 in that for each, there exists an algorithm
that, given a finite structure, tells what the complexity of the corresponding
problem is.
For the family of problems $\csp(\A)$,
Feder and Vardi~\cite{FederVardi99-structure}
famously conjectured that
there is a dichotomy in the finite:
for each finite structure $\A$, the problem $\csp(\A)$ is either
polynomial-time tractable or NP-complete.
Investigation of the complexity-theoretic properties
of the problem families $\csp(\A)$ and $\qcsp(\A)$, on finite structures,
is a research theme of
active interest~\cite{Chen08-collapsibility,ABISV09-refining,LaroseTesson09-hardness,BartoKozik09-boundedwidth,BBCJK09-qcsp,IMMVW10-tractabilityfewsubpowers,Chen11-qcspandpgp,Chen12-meditations}.

At the heart of the work on these classification programs are
\emph{algebraic preservation theorems} which
state that, relative to a finite structure,
the relations definable in a given fragment are precisely those
preserved by a suitable set of operations.
As an example, one such theorem
states that a relation is primitive positive definable on a finite
structure $\A$ if and only if all polymorphisms of $\A$ are
polymorphisms of the
relation~\cite{Geiger68-closed,BKKR69-galois}.
(A polymorphism of a
structure $\A$ is a homomorphism from a finite power $\A^k$ to $\A$
itself.) On finite structures there are analogous preservation
theorems connecting positive Horn definability to surjective
polymorphisms~\cite{BBCJK09-qcsp},
existential positive definability to
endomorphisms~\cite{Krasner68-endomorphisms},
and equality-free positive definability to
so-called surjective hyper-endomorphisms~\cite{MadelaineMartin09-complexityPFOwithouteq}.
For the purposes
of complexity classification, these preservation theorems are
relevant in that they allow one to pass from the study of structures
to the study of algebraic objects.
For instance, it follows from the preservation theorem for primitive
positive definability that two finite structures $\A, \B$ having the
same polymorphisms are primitive positively interdefinable, from which
it readily follows that the problems $\csp(\A)$ and $\csp(\B)$ are
interreducible and share the same complexity
(under many-one logspace reduction); thus, insofar as one is
interested in CSP complexity, one can focus on investigating the
polymorphisms of structures.

Given the import and reach of these algebraic preservation theorems
for finite structures, a natural consideration is to generalize them to
infinite structures. Although it is known that these preservation
theorems do not hold on \emph{all}
infinite structures (see the discussion in~\cite{BodirskyHilsMartin10-scopeunivalg}
as well as~\cite[Theorem 4.7]{Bodirsky04-thesis}),
Bodirsky and Ne\v{s}et\v{r}il \cite[Theorem~5.1]{BodirskyNesetril06-homogeneous}
established that the
preservation theorem characterizing primitive positive definability
via polymorphisms
does hold on
$\aleph_0$-categorical structures, which have countably infinite
universes.
An $\aleph_0$-categorical structure is ``finite-like'' in that
for each fixed arity,
there are a finite number of first-order definable relations;
indeed, this is one of the characterizations of
$\aleph_0$-categoricity given by the classical theorem of
Ryll-Nardzewski.
The class of $\aleph_0$-categorical structures
includes many structures of computational interest,
including those whose relations are first-order definable
over one of the following structures:
equality on a
countable universe,
  the ordered rationals $(\mathbb Q, <)$, and
  the countable random graph;
see \cite{bodsurvey} for a survey.

In this paper, we present
an algebraic preservation theorem for positive
Horn definability on $\aleph_0$-categorical structures. This theorem
characterizes positive Horn definability by making use of a
construction
which we call the \emph{periodic power}. In particular, we define a
\emph{periomorphism} of a structure $\A$ as a homomorphism from the
periodic power of $\A$ to $\A$ itself, and show that a relation is
positive Horn definable over an $\aleph_0$-categorical structure
$\A$ if and only if all surjective periomorphisms of $\A$ are
periomorphisms of the relation.

The periodic power of a structure $\A$ is the substructure of
$\A^{\mathbb N}$ whose universe is the set of all periodic tuples in
$\A^{\mathbb N}$; a tuple $(a_0, a_1, \ldots)$
is periodic if
there exists an integer $k \geq 1$ such that the tuple
\emph{repeats mod $k$},
by which is meant $a_n = a_{n \Mod k}$ for all $n \in \mathbb N$.
As we discuss in the paper, the periodic power arises as the
direct limit of an appropriately defined system of embeddings.
Despite the extremely natural character of the periodic power, we
are not aware of previous work where this construction has been
explicitly considered.
We believe that it could be worthwhile to seek applications of the
periodic power in other areas of mathematics.
One basic fact
that we demonstrate is that the positive Horn theory of a structure
holds in the structure's periodic power; this readily implies that
the class of groups is closed under 
periodic powers,
and likewise 
for other classes
of classical algebraic structures such as rings, lattices, and Boolean algebras.
Our introduction and study of the periodic power also forms a contribution of this paper.

A direct corollary of our preservation theorem is that for two $\aleph_0$-categorical structures
$\A,\B$ with the same universe, 
if $\A$ and $\B$ have the same surjective periomorphisms,
then the structures $\A$ and $\B$ are positive Horn interdefinable,
and the computational problems $\qcsp(\A)$ and $\qcsp(\B)$ are interreducible
(under many-one logspace reductions).
This permits the use of surjective periomorphisms in the study
of the complexity of the QCSP on $\aleph_0$-categorical structures.
As an application of our preservation theorem
and the associated theory that we develop,
we give a new proof of
the known complexity classification of \emph{equality templates},
which are structures whose relations are first-order definable
over the equality relation on a countable set.

\subsection*{Related work} 
An algebraic preservation theorem for positive Horn definability via
surjective polymorphisms was shown for the special case of equality templates~\cite{BodirskyChen10-equality}.
The presented proof crucially depends on results on the clones of equality templates
given there and in~\cite{BodirskyChenPinsker10-reducts}.

In model theory, there are \emph{classical preservation theorems}
that show that a sentence is equivalent to one in a given fragment
if and only if its model class satisfies some suitable closure properties.
Such theorems have been shown for positive Horn logic.
A well-known instance is
Birkhoff's HSP theorem characterizing universally quantified equations.
And in 1955, Bing~\cite{bing} showed that
a positive sentence is preserved by direct products if and only if
it is equivalent to a positive Horn sentence. Later,
assuming the continuum hypothesis (CH),
Keisler proved\footnote{In fact, Keisler could do assuming
only the existence of some cardinal $\kappa\ge\aleph_0$ such that $2^{\kappa}=\kappa^+$.}
 that a sentence is equivalent to a positive Horn sentence if and only if it
is preserved  (in the parlance of~\cite{vogler,flum}) by the following binary relation:
relate $\A$ to $\B$ when $\B$ is a homomorphic image of $\A^{\mathbb N}$~\cite[Corollary~3.8]{a}
(see also~\cite[Section 6.2]{keisler}). Absoluteness considerations can be used to eliminate
the assumption of CH when one has ZFC provability of the stated closure property.
More recently,  Madelaine and Martin~\cite[Theorem~1]{mm}  showed, without relying on CH, that Keisler's result holds
when one considers preservation under the relation defined as above, but where $\B$ is required to be finite.

In some cases, an algebraic preservation theorem can be derived
from a corresponding classical preservation theorem.
Such a derivation has been given
for Bodirsky and Nesetril's theorem in~\cite{bodsurvey},
and Bodirsky and Junker~\cite{junker}
derived algebraic preservation
theorems for existential positive definability and positive definability in
$\aleph_0$-categorical structures from well-known
classical preservation theorems of Lyndon.
%
%
%
%
%
Roughly speaking,
these methods need the preservation relation to be $\textit{PC}_\Delta$ (cf.~\cite{flum} or \cite[p.103]{vogler}) and thus
cannot be applied to Keisler's classical preservation theorem mentioned above. To the
best of our knowledge, prior to this work no algebraic preservation theorem for positive Horn formulas on $\aleph_0$-categorical structures
has been known (neither in the presence nor absence of~CH).
%
%

\section{Preliminaries from model theory}

\subsection{First-order logic} Throughout the paper, $L$ will denote a countable first-order language.
If not explicitly stated otherwise, by a structure (formula) we always mean an $L$-structure
(first-order $L$-formula). Throughout, we use the letters $\A$, $\B$, etc.
to denote structures with universes $A, B$, etc.; 
we use $\varphi,\psi,\chi,$ etc. to denote formulas.
For a  structure $\A$ and a (finite) tuple $\bar a$ from $A$, by $(\A,\bar a)$ we denote, as usual, the expansion of $\A$ interpreting
new constants by the components of $\bar a$. We do not distinguish between constants outside $L$ and variables.
For a formula $\varphi=\varphi(\bar x)$ and a structure $\A$, writing $(\A,\bar a)\models\varphi(\bar x)$ or $\A\models\varphi(\bar a)$ (with $\bar x$ clear from context)
means that $\A$ satisfies $\varphi(\bar x)$ under the assignment $\bar a$ to $\bar x$. By $\varphi(\A)$ we denote the relation
$\{\bar a\mid \A\models\varphi(\bar a)\}$ on $A$; this relation is said to be {\em defined by $\varphi$ in $\A$}.
A relation is
{\em first-order (positive Horn, primitive positively) definable in $\A$}
if it is defined by some first-order (positive Horn, primitive positive) formula
$\varphi$ in $\A$ (see Section~\ref{subsec:ph} for definitions of positive Horn and primitive positive).

Let $L'$ be another first-order language, $\B$ an $L'$-structure and $\A$ an $L$-structure such that $A=B$. Then
$\B$ is {\em first-order (positive Horn, primitive positively) definable in $\A$} if for every 
atomic $L'$-formula $\varphi$ the relation $\varphi(\B)$ is (positive Horn, primitive positively) definable in $\A$.

\subsection{Direct products}
For a family of ($L$-)structures we denote its direct product by $\prod_{i\in I}\A_i$. Recall that this structure
\begin{enumerate}[--]

\item has universe $\prod_{i\in I}A_i$,
which is the set of functions mapping each $i\in I$ into the universe $A_i$ of $\A_i$;

\item interprets a $k$-ary relation symbol $R\in L$ by those $k$-tuples $(\vec a_0,\ldots, \vec a_{k-1})$ from $\prod_{i\in I} A_i$
such that $\A_i\models R\vec a_0(i)\cdots\vec a_{k-1}(i)$ for all $i\in I$; and

\item interprets a $k$-ary function symbol $f\in L$ by the function mapping a $k$-tuple $(\vec a_0,\ldots, \vec a_{k-1})$ from
$\prod_{i\in I} A_i$ to the element $\vec a\in\prod_{i\in I}A_i$
having the property that $\A_i\models f(\vec a_0(i),\ldots, \vec a_{k-1}(i))=\vec a(i)$ for all $i\in I$.
\end{enumerate}
We write $\A^I$ for $\prod_{i\in I}\A_i$ with all $\A_i=\A$;
we write $\A^k$ to indicate $\A^I$
when $I=\{0,\ldots,k-1\}$ for $k\in\mathbb N,k>0$.
We consider $\A^k$ to have universe $A^k$, the set of $k$-tuples over $A$. We do not distinguish between 1-tuples
and elements, that is,
 $\A^1=\A$.
The direct product of two structures $\A$ and $\B$ is denoted $\A\times \B$ and considered to have universe $A\times B$.

\subsection{Direct limits} 
We recall the definitions associated with direct limits. 
Let $(I,\prec)$ be a strict partial order that is directed:
every two elements in $I$ have a common upper bound. An {\em $(I,\prec)$-system of embeddings (homomorphisms)}
is a family of embeddings (homomorphisms) $e_{(i,j)}:\A_i\to\A_j$ for $i\prec j$ such that
$e_{(i,k)}=e_{(j,k)}\circ e_{(i,j)}$ for all $i\prec j\prec k$. A {\em cone} of the system is a family of {\em limit embeddings (homomorphisms)}
$e^*_i:\A_i\to\A^*$ such that  $e^*_j\circ e_{(i,j)}=e^*_{i}$.
It is known that, for a system, there exists a cone  satisfying the following universal property:
for every other cone, say given by $\tilde\A $ and $ (\tilde e_i)_{i\in I}$, there exists a unique embedding (homomorphism) $e:\A^*\to\tilde\A$ such that
$e\circ e^*_i=\tilde e_i$. A structure $\A^*$
with this universal property is unique up to isomorphism and called the {\em direct limit} of the system;
if $(I,\prec)$ and the $e_{(i,j)}$s are clear from context, 
it is denoted by $\lim_i \A_i$.

\subsection{$\aleph_0$-categoricity}\label{subsec:cat}
A structure $\mathfrak A$ is {\em $\aleph_0$-categorical} if it is countable and every countable structure $\B$ that satisfies
the same first-order sentences as $\mathfrak A$ is isomorphic to $\mathfrak A$. We assume basic familiarity with $\aleph_0$-categoricity
as covered by any standard course in model theory 
(see for example~\cite{keisler}).
Here, we briefly recall some facts that we are going to use.

The theorem of {\em Ryll-Nardzewski} states that a countable structure $\mathfrak A$ is $\aleph_0$-categorical if and only if for every $k\in\mathbb N$ there are at most finitely many $k$-ary relations that are
first-order definable in $\A$. It is straightforward to verify
 that this implies that for an $\aleph_0$-categorical
structure $\A$, when $\bar a$ is an arbitrary finite-length tuple
from $A$, the structure
$(\A,\bar a)$ is also $\aleph_0$-categorical.
Further, it implies that
for an $\aleph_0$-categorical structure $\A$,
the structure $\A^k$ is $\aleph_0$-categorical for any
$k\in\mathbb N$; in fact,
every structure that is first-order interpretable in an $\aleph_0$-categorical structure is also $\aleph_0$-categorical.

Another easy consequence
of this theorem is that $\aleph_0$-categorical structures are {\em $\aleph_0$-saturated},
by which is meant that for every finite tuple $\bar a$ from $A$ and
every set of formulas $\Phi=\Phi(x)$ in the language of $(\A,\bar a)$
(that is, having constants for~$\bar a$) one has: if every finite subset of
$\Phi(x)$ is  satisfiable in $(\A,\bar a)$, then so is $(\A,\bar a)$.

Finally, we mention the fact that for an $\aleph_0$-categorical structure $\A$,
a relation over $A$ is first-order definable if and only if it is
preserved by all automorphisms of $\A$ (see Section~\ref{subsec:pres} for the definition of preservation).

\section{Preliminaries from constraint satisfaction} \label{subsec:ph}

\subsection{Positive Horn formulas} As noted in the introduction, a {\em positive Horn} formula is a
first-order formula built from atoms, conjunction, and the two quantifiers.
Existential such formulas are {\em primitive positive}.
For simplicity, we assume that first-order logic contains a propositional constant $\bot$ for falsehood;
formally, $\bot$ is a 0-ary relation symbol always interpreted by $\emptyset$.
Note that $\bot$ is a positive atomic sentence. If any positive Horn sentence true in $\A$ is also true in $\B$, we write
$\A\Rrightarrow_{\textup{pH}}\B$.

A formula $\varphi(\bar x)$ is {\em preserved by direct products} if it holds in $(\A,\bar a)\times(\B,\bar b)$ whenever it holds in
both $(\A,\bar a)$ and $(\B,\bar b)$. Positive Horn formulas are preserved by direct products, in fact,
the following is straightforward to verify.

\begin{lem}\label{lem:basic} Let $(\A_i)_{i\in I}$ be a family of structures. A positive Horn sentence holds in $\prod_{i\in I}\A_i$ if and only if it
holds in every~$\A_i,i \in I$.
\qed \end{lem}

\subsection{Quantified constraints} The quantified constraint satisfaction problem (QCSP) on a structure $\A$, denoted by $\qcsp(\A)$, is the problem of deciding the positive Horn
theory of $\A$. The following proposition relates positive Horn definability to the complexity of the QCSP.
 
\begin{prop}
\label{prop:qcsp-reduction} Let $\A$ be an $L$-structure and $\B$ be an $L_0$-structure for some finite first-order language $L_0$.
If $\B$ is positive Horn definable in $\A$, then the problem $\qcsp(\B)$ many-one logspace reduces to $\qcsp(\A)$.
\end{prop}

\proof For every function symbol $f\in L_0$, constant $c\in L_0$ and relation symbol $R\in L_0$ choose some fixed positive
Horn $L$-formulas $\psi_f(\bar x,y), \psi_c(x), \psi_R(\bar x)$ that respectively define, in $\A$,
the relations given by the formulas
$f(\bar x)=y, x=c,R\bar x$ interpreted over $\B$.
Let~$\varphi$ be an instance of $\qcsp(\B)$, that is, a positive Horn sentence in the language~$L_0$.
In a first step, compute in logspace
an equivalent sentence $\varphi^*$ in which  every atomic subformula
 contains at most one symbol from $L_0$,
that is, has the form $x=y,f(\bar x)=y$ or $R\bar x$.
This can be done by successively replacing atomic subformulas of $\varphi$,
for example, replacing $Rxcf(f(x))$ by
$$
\exists y_0y_1y_2(Rxy_0y_2\wedge y_0=c\wedge f(y_1)=y_2\wedge f(x)=y_1)
$$
In a second step, replace in $\varphi^*$ every atomic subformula that mentions $s\in L_0$ by the formula $\psi_s$
(with the right choice of variables). This can also be done in logspace: note that we may hardwire the finite list of the
formulas $\psi_s$ into the algorithm.
Finally, recall that the composition of two logspace  algorithms can be implemented in logspace.
\proofend

\begin{rem}
In the literature, the CSP and QCSP are typically defined in \emph{relational}
first-order logic. We take a more general stance and allow the language to contain function symbols if not explicitly stated otherwise.
In particular, our preservation theorem
(Theorem~\ref{theo:pres})
 holds in the presence of function symbols.
\end{rem}

\subsection{Preservation}   \label{subsec:pres}
Let $A$ be a set, $I$ a nonempty set and $h$ a partial function from $A^I$ to $A$.
If $h$ is defined on all of $A^I$ (and $I$ is finite),
it is called a {\em (finitary) operation} on $A$.
Then $h$ is said to {\em preserve} an $r$-ary relation
$R\subseteq A^r$ if it is a partial homomorphism from $(A,R)^I$ to $(A,R)$. This means the following: whenever
$\vec a_0,\ldots,\vec a_{r-1}$ are in the domain of $h$ and $(\vec a_0(i),\ldots,\vec a_{r-1}(i))\in R$ for
all $i\in I$, then $(h(\vec a_0),\ldots,h(\vec a_{r-1}))\in R$.
Further, 
relative to a structure $\A$ with universe $A$, 
we say that $h$ preserves a formula~$\varphi$
 if it preserves the relation
$\varphi(\A)$.

\subsection{Clones and Polymorphisms} \label{subsec:poly}
A {\em clone on $A$} is a set of finitary operations on $A$ that is closed under composition and contains all projections.
A set $F$ of operations on $A$ {\em interpolates}
an operation $g$ on $A$ if for all finite sets $B$ there exists an operation $f\in F$ such that
$f\upharpoonright B=g\upharpoonright B$.
A set of operations is {\em locally closed}
if it contains every operation that it interpolates.

A {\em polymorphism of $\A$} is a homomorphism from $\A^k$ to $\A$ 
where $k$ is a positive integer
called the {\em arity} of the polymorphism.
Equivalently,
a polymorphism of $\A$ is a finitary operation on $A$ that
preserves each $\A$-relation, $\A$-constant, and graph of an $\A$-function;
or, a polymorphism of $\A$ is a finitary operation on $A$ that preserves all atomic formulas.
It is straightforward to verify  that the set of polymorphisms of 
any structure $\A$ forms a
locally closed clone on~$A$.

An operation $h: A^k \rightarrow A$ is a polymorphism of a relation
$R \subseteq A^{\ell}$
if $h$ is a polymorphism of the structure $(A, R)$.
In a picture, this means the following.
If every column of
$$
\begin{array}{llll}
a_{0}^0&a_{1}^0&\cdots&a_{k-1}^0\\
a_{0}^1&a_{1}^1&\cdots&a_{k-1}^1\\
\ \ \vdots&\ \vdots&\ddots&\ \vdots\\
a_{0}^{\ell-1}&a_{1}^{\ell-1}&\cdots&a_{k-1}^{\ell-1}
\end{array}
$$
is a tuple contained in $R$,
then so is the $\ell$-tuple obtained by applying $h$ to each row.

We have the following polymorphism-based characterization
of primitive positive definability.
\begin{thm}[\cite{BodirskyNesetril06-homogeneous}]
\label{theo:inv-pol}
Let $\A$ be $\aleph_0$-categorical. A relation $R$ over $A$ is primitive positively definable in $\A$
if and only if it is preserved by all polymorphisms of~$\A$.
\qed \end{thm}


\section{Periodic powers}\label{sec:periodicpower}

In this section, we present the notion of the \emph{periodic power} of
a structure, and identify some basic properties thereof.
We also discuss how the periodic power arises as the direct limit
of a system of embeddings.
Throughout this section, we use $\A,\B$ to denote structures.

\begin{defi} A function $\vec a:\mathbb N\to A$ is {\em periodic}
if there exists $k\in\mathbb N, k > 0$
such that for all $i \in \mathbb N$,
it holds that $\vec a(i)=\vec a(i\Mod k)$;
in this case
the function $\vec a$ is said to be {\em $k$-periodic},
and we write $\langle \vec a(0)\cdots\vec a(k-1)\rangle$ to denote $\vec a$.
The set of periodic functions $A^{\textup{per}}$ 
carries a substructure in $\A^{\mathbb N}$:
the set $A^{\textup{per}}$
is nonempty and  closed under all $\A^{\mathbb N}$-interpretations
of function symbols.
We define the {\em periodic power} of $\A$, denoted $\A^{\textup{per}}$,
to be the substructure of $\A^{\mathbb N}$ induced on $A^{\textup{per}}$.
\end{defi}

When $\bar{\vec a}=\vec a_0\cdots\vec a_{\ell-1}$ is a tuple from $A^{\textup{per}}$ and $i\in\mathbb N$, we let 
$\bar{\vec a}(i)$ denote the tuple 
$\vec a_{0}(i)\cdots\vec a_{\ell-1}(i)$ from $A$.

\begin{lem}\label{lem:basic2}
Assume that $\varphi(\bar x)$ is a positive Horn formula.
Then $(\A^{\textup{per}},\bar{\vec a})\models\varphi(\bar x)$ if and only if
$(\A,\bar{\vec a}(i))\models\varphi(\bar x)$ for all $i\in\mathbb N$.
\end{lem}

\proof
Call a formula $\varphi$ {\em good} if it satisfies the claimed equivalence. Clearly, conjunctions of atoms are good.
Assume $\varphi(\bar x,y)$ is good. It is easy to see that also $\forall y\varphi(\bar x,y)$ is good. We show that $\exists y\varphi(\bar x,y)$ is  good,
via the following equivalences.
\begin{align}\nonumber
&(\A^{\textup{per}},\bar{\vec a})\models\exists y\varphi(\bar x,y)\\\nonumber
&\Longleftrightarrow
\exists\vec b\in A^{\textup{per}}: \ (\A^{\textup{per}},\bar{\vec a},\vec b)\models\varphi(\bar x,y)
\\\label{eq:2}
&\Longleftrightarrow \exists\vec b\in A^{\textup{per}} \ \forall i\in\mathbb N: \ (\A,\bar{\vec a}(i),\vec b(i))\models\varphi(\bar x,y)
\\\label{eq:3}
&\Longleftrightarrow \forall i\in\mathbb N \ \exists b\in A: \ (\A,\bar{\vec a}(i), b)\models\varphi(\bar x,y)
\\\nonumber
&\Longleftrightarrow  \forall i\in\mathbb N : \ (\A,\bar{\vec a}(i))\models\exists y \varphi(\bar x,y).
\end{align}

\noindent The second equivalence follows from $\varphi(\bar x,y)$ being good. The rest being trivial,
we show  that \eqref{eq:3} implies \eqref{eq:2}.
By \eqref{eq:3} there is a function $\vec b:\mathbb N\to A$ such that $(\A,\bar{\vec a}(i),\vec b(i))\models\varphi(\bar x,y)$ for all $i\in\mathbb N$. For every component $\vec a$ of $\bar{\vec a}$ choose $n_{\vec a}\in\mathbb N$ such that $\vec a$ is $n_{\vec a}$-periodic, and let $n\in\mathbb N$ be a common multiple of the $n_{\vec a}$s. Then any component of $\bar{\vec a}$ is $n$-periodic and, in particular,
$$
\bar{\vec a}(i)=\bar{\vec a}(i\Mod n)
$$
for all $i\in\mathbb N$. Define $\vec b^*:\mathbb N\to A$ by
$$
\vec b^*(i):=\vec b(i \Mod n).
$$
Then $\vec b^*\in A^{\textup{per}}$ and $(\A,\bar{\vec a}(i), \vec b^*(i))\models\varphi(\bar x,y)$ for all $i\in\mathbb N$; this is \eqref{eq:2}.
\proofend


Consider the following embeddings.
\begin{enumerate}[--]

\item The function $e_1:\A\to\A^{\textup{per}}$ defined by
$e_1(a):=\langle a\rangle$, that is, the function mapping each
$a \in A$ to the constant sequence $(a)_{i\in\mathbb N}$,
is a canonical embedding of $\A$ into $\A^{\textup{per}}$.

\item More generally, for each  $k > 0$,  the function $e_k:\A^k\to\A^{\textup{per}}$
defined by $e_k((a_0,\ldots,a_{k-1})):=\langle a_0\cdots a_{k-1}\rangle$
is a canonical embedding from $\A^k$ into $\A^{\textup{per}}$.

\end{enumerate}
In the following proposition we identify $a\in A$ with $e_1(a)\in A^{\textup{per}}$ for notational simplicity.
We use $\A\preceq_{\textup{pH}}\B$ to indicate that
$\A\subseteq \B$ (i.e. $\A$ is a substructure of $\B$)
 and that
for every positive Horn formula $\varphi(\bar x)$ and all tuples $\bar a$ from $A$, it holds that
$$
(\A,\bar a)\models\varphi(\bar x)\Longleftrightarrow(\B,\bar a)\models\varphi(\bar x).
$$
 Lemmas~\ref{lem:basic} and~\ref{lem:basic2} imply:

\begin{prop}\label{prop:el} $\A\preceq_{\textup{pH}} \A^{\textup{per}}\preceq_{\textup{pH}}  \A^{\mathbb N}$.
\qed \end{prop}

The next two propositions explain how the periodic power relates to finite powers.

\begin{prop}\label{prop:iso} Let $k\in\mathbb N,k>0$. Then $\A^{\textup{per}}\cong (\A^k)^{\textup{per}}$ via an isomorphism that
maps $\langle a_0\cdots a_{k-1}\rangle$ to $\langle (a_0,\ldots,a_{k-1})\rangle$ for all $a_0,\ldots,a_{k-1}\in A$.
\end{prop}

To make clear the notation used in the statement of this proposition, let us look at an example:
the notation $\langle ab \rangle$ denotes the 2-periodic sequence $ababab\cdots\in A^{\textup{per}}$,
whereas the notation
$\langle (a,b) \rangle$ denotes the constant, 1-periodic
sequence $(a,b)\ (a,b)\ (a,b)\cdots\in (A^2)^{\textup{per}}$.\medskip

\noindent{\em Proof of Proposition~\ref{prop:iso}.} 
Define the map
$f:\A^{\textup{per}}\to(\A^k)^{\textup{per}}$
to map $\vec a\in \A^{\textup{per}}$ to
$$
i\mapsto (\vec a(ik),\ldots,\vec a((i +1)k-1))
$$
The map $f$ is clearly injective.  For $j<k$ let $\pi^k_j$ denote the projection of $k$-tuples to their $(j+1)$th component.
An element $\vec b\in(\A^k)^{\textup{per}}$ has
$$
i\mapsto \pi^k_{i\Mod k} (\vec b(\lfloor i/k\rfloor) )
$$
as preimage under $f$, so $f$ is surjective.  It is straightforward to verify that $f$ is an isomorphism.
\proofend

\begin{prop} \label{prop:iso2} Let $k\in\mathbb N,k>1$. Then $\A^{\textup{per}}\cong (\A^{\textup{per}})^k$.
\end{prop}

The proof relies on the following observation.

\begin{lem}\label{lem:prod} $\A^{\textup{per}}\times \B^{\textup{per}}\cong (\A\times\B)^{\textup{per}}$.
\end{lem}
\proof Map a pair of functions $(\vec a,\vec b)\in A^{\textup{per}}\times B^{\textup{per}}$ to  $((\vec a(i),\vec b(i)))_{i\in\mathbb N}$; note this
function is $nm$-periodic whenever $\vec a$ and $\vec b$ are $n$- and $m$-periodic respectively.
The map is clearly injective. It is surjective as
  $((a_i,b_i))_{i\in\mathbb N}\in (A\times B)^{\textup{per}}$ has preimage
$((a_i)_{i\in \mathbb N},(b_i)_{i\in\mathbb N})\in A^{\textup{per}}\times B^{\textup{per}}$.
%
To see that it is an isomorphism,
let $\alpha$ be an atom. For simplicity assume $\alpha=\alpha(x,y)$, and
let $(\vec a,\vec b),(\vec a',\vec b')\in A^{\textup{per}}\times B^{\textup{per}}$. Then
\begin{eqnarray*}
&&(\A^{\textup{per}}\times \B^{\textup{per}},(\vec a,\vec b),(\vec a',\vec b'))\models\alpha(x,y) \\
&&\Longleftrightarrow
(\A^{\textup{per}},\vec a,\vec a')\models\alpha(x,y)\textup{ and } ( \B^{\textup{per}},\vec b,\vec b')\models\alpha(x,y)
\\
&&\Longleftrightarrow \forall i\in\mathbb N:\ (\A,\vec a(i),\vec a'(i))\models\alpha(x,y)
\textup{ and } ( \B,\vec b(i),\vec b'(i))\models\alpha(x,y)
\\
&&\Longleftrightarrow \forall i\in\mathbb N :\ (\A\times\B, (\vec a(i),\vec b(i)),(\vec a'(i),\vec b'(i)))\models\alpha(x,y)
\\
&&\Longleftrightarrow ((\A\times\B)^{\textup{per}}, (\vec a(i),\vec b(i))_{i\in\mathbb N},  ((\vec a'(i),\vec b'(i)))_{i\in\mathbb N})\models\alpha(x,y),
\end{eqnarray*}
where the first and third equivalence hold by definition of direct products, and the second and fourth equivalence hold by Lemma~\ref{lem:basic2}.
\proofend

\noindent{\em Proof of Proposition~\ref{prop:iso2}} by induction on $k$:
we have the isomorphisms
$$
(\A^{\textup{per}})^{k+1}=(\A^{\textup{per}})^k\times \A^{\textup{per}} \cong \A^{\textup{per}}\times \A^{\textup{per}} \cong (\A^2)^{\textup{per}}
\cong \A^{\textup{per}}
$$
by induction, the previous lemma and Proposition~\ref{prop:iso}.\proofend

Observe that for $n,m>0$ there is a natural embedding $e_{(n,m)}:\A^n\to\A^{m}$ whenever $n<m$ and $n$ divides $m$, namely the embedding that maps the $n$-tuple $\bar a\in A^n$ to the $m$-tuple
$$
e_{(n,m)}(\bar a)=\underbrace{\bar a\bar a\cdots\bar a}_{m/n\text{ times}}\in A^m.
$$
Clearly, these embeddings are compatible
in the sense that $e_{(\ell,m)}\circ e_{(n,\ell)}=e_{(n,m)}$ whenever $n<\ell<m$, $n$ divides $\ell$ and $\ell$ divides $m$. In other words,
the $e_{(n,m)}$s determine an $(I,\prec)$-system of embeddings where $I=\mathbb N_{>0}$ and  $\A_n:=\A^n$ and $\prec $ denotes divisibility. 

\begin{prop}\label{prop:lim} $\A^{\textup{per}}\cong \lim_n \A^{n}$.
\end{prop}

\proof Let $(e^*_n)_{n>0}$ denote the limit homomorphisms into the direct limit $\lim_n \A^n$ of the directed system of embeddings given by the 
embeddings $e_{(n,m)}$ (for $n<m$ and $n$ divides $m$).
Observe that the embeddings $e_n$ from $\A^n$ into $\A^{\textup{per}}$ satisfy the requirement for limit embeddings, so these embeddings $e_n$
 are also a cone of the directed system.
By the universal property of  $\lim_n \A^n$ there is an
embedding $e:\lim_n \A^n\to\A^{\textup{per}}$ such that $e\circ e^*_n=e_n$ for all $n>0$.
But every element of $ A^{\textup{per}}$ is
in the image of some~$e_n$, so $e$ has to be surjective and thus is an isomorphism. \proofend

Recall, an $\forall\exists$-sentence is a sentence of the form $\forall\bar x\exists \bar y\psi$ with $\psi$ quantifier free.
Propositions~\ref{prop:el} and~\ref{prop:lim} imply:

\begin{cor}
Every positive Horn sentence true in $\A$ and every $\forall\exists$-sentence
 true in all finite powers of $\A$, is true in~$\A^{\textup{per}}$.
\qed \end{cor}


\section{Periomorphisms}\label{sec:periomorphisms}

In this section, we introduce and study the notion of \emph{periomorphism}.
Throughout this section, let $\A$ be a structure.

\begin{defi} A {\em periomorphism of $\A$} is a homomorphism from $\A^{\textup{per}}$ to $\A$.
\end{defi}

In other words, a periomorphism of $\A$ is a partial
function from $A^{\mathbb N}$ to $A$ with domain $A^{\textup{per}}$ that preserves all atomic formulas.
The following lemma follows straightforwardly from the definitions.

\begin{lem} \label{lem:boff}
A periomorphism $h$ of $\A$ preserves a
relation $R \subseteq A^{\ell}$
if and only if for any choice of finitely many tuples
$\bar a_0= (a_{0}^0,\ldots, a_{0}^{\ell-1}),\ldots, \bar a_{k-1}=(a_{k-1}^0,\ldots, a_{k-1}^{\ell-1})$ from $R$, we have
$$
\big(h(\langle a_{0}^0 a_{1}^0 \cdots a_{k-1}^0\rangle),\ldots,h(\langle a_{0}^{\ell-1} a_{1}^{\ell-1} \cdots a_{k-1}^{\ell-1}\rangle)\big) \in R.
$$
\end{lem}

\noindent{\em Proof:}
The forward direction is trivial. Conversely assume the right hand side of the claimed equivalence and let
$\vec a_0,\ldots,\vec a_{\ell-1}\in \A^{\textup{per}}$
be such that  for all $i\in\mathbb N$, $(\vec a_0(i),\ldots \vec a_{\ell-1}(i))\in R$.
We claim
$h(\vec a_0)\cdots h(\vec a_{\ell-1})\in R$. Choose a sufficiently large $k\in\mathbb N$ such
that all $\vec a_j$ are $k$-periodic, that is, $\vec a_{j}=\langle\vec a_j(0)\cdots \vec a_j(k-1)\rangle$ for all $j<\ell$.
Applying the assumption yields the claim.
\proofend

To see the lemma's statement with a picture, let
$h$ be a periomorphism of $\A$, and consider the following.
$$
\begin{array}{llll}
\langle a_{0}^0&a_{1}^0&\cdots&a_{k-1}^0\rangle\\
\langle a_{0}^1&a_{1}^1&\cdots&a_{k-1}^1\rangle\\
\ \ \vdots&\ \vdots&\ddots&\ \vdots\\
\langle a_{0}^{\ell-1}&a_{1}^{\ell-1}&\cdots&a_{k-1}^{\ell-1}\rangle
\end{array}
$$
The right hand side of the lemma states that if the columns $\bar a_i= (a_{i}^0,\ldots,
a_{i}^{\ell-1})$ are contained in
$R$ for all $i < k$, then so is the
$\ell$-tuple $\bar b$ obtained by applying $h$ to each
row.

For later use we introduce the following mode of speech.

\begin{defi}\label{defn:image} In the situation above,
if $h$ is a \emph{surjective} periomorphism of the structure under study,
then we call $\bar b$
{\em a surjective periomorphic image of the tuples $(\bar a_i)_{i<k}$}.
\end{defi}

\begin{prop}\label{prop:pres}
Every positive Horn formula is preserved by
 all surjective periomorphisms of $\A$.
\end{prop}

\noindent{\em Proof:} Let $\varphi(\bar x)$ be a positive Horn formula and $h$ be a surjective periomorphism of $\A$.
For notational simplicity assume $\bar x=xx'$ and let $a_0a_0', \ldots, a_{k-1}a_{k-1}'$ be any finitely
 many pairs in $\varphi(\A)$. We have to show that $\varphi(xx')$ is true in $(\A,h(\langle a_0\cdots a_{k-1}\rangle),h(\langle a_0'\cdots a'_{k-1}\rangle))$;
see the previous lemma.
But $\varphi(xx')$ is true in $(\A^{\textup{per}},\langle a_0\cdots a_{k-1}\rangle,\langle a_0'\cdots a'_{k-1}\rangle)$ by
 Lemma~\ref{lem:basic2}, and, being positive, is preserved by surjective homomorphisms.\proofend

The periomorphisms and the polymorphisms of a structure contain
the same information.
If one knows the periomorphisms of a structure, then one also knows
its polymorphisms -- and vice-versa.  Why is this?
For $k\in\mathbb N,k>0$ define
$\pi_{<k}:A^{\textup{per}} \to A^k$ by
$$
\pi_{<k}(\vec a):= (\vec a(0),\ldots,\vec a(k-1)).
$$
This operation is clearly a homomorphism from
$\A^{\textup{per}}$ to $\A^k$.
Now, if someone hands us an operation $h: A^k \to A$,
we can decide if it is a polymorphism of $\A$ by checking
if
$$
h^{\textup{per}}:=h\circ\pi_{<k}.
$$
is a periomorphism of $\A$.
For,
if $h$ is a polymorphism of $\A$, then by composing homomorphisms,
we have that
$h^{\textup{per}}$ is a periomorphism of $\A$;
and, if
$h^{\textup{per}}$ is a periomorphism of $\A$,
by composing homomorphisms,
we have that
$h^{\textup{per}} \circ e_k$, which is equal to $h$,
 is a homomorphism from $\A^k$ to $\A$.

Going the other way, suppose that someone places in our hands an operation
$h: A^{\textup{per}} \to A$.
It can be seen from Lemma~\ref{lem:boff}
that $h$ is a periomorphism
of $\A$ if and only if each of the operations
\begin{equation}\label{eq:fk}
h_{<k}:= h\circ e_k.
\end{equation}
is a polymorphism of $\A$.


It is thus no surprise that preservation by periomorphisms coincides with preservation by polymorphisms. Preservation by {\em surjective} periomorphisms,
however,
is an a priori stronger property than preservation by surjective polymorphisms.

\begin{prop}\label{prop:polyper} Let $\varphi$ be a formula. Then
\begin{enumerate}[\em(1)]
\item $\varphi$ is preserved by all periomorphisms of $\A$ if and only if $\varphi$  is preserved by all polymorphisms of $\A$;
\item if $\varphi$ is preserved by all surjective periomorphisms of $\A$, then $\varphi$  is preserved by all surjective polymorphisms of~$\A$.
\end{enumerate}
\end{prop}

\proof To see the forward directions, observe that if $h$ is a (surjective) polymorphism of $\A$ that does not preserve $\varphi$, then $h^{\textup{per}}$ is a (surjective) periomorphism of $\A$ that does not preserve $\varphi$.
For the converse direction in (1) assume $h$ is a periomorphism that does not preserve
$\varphi=\varphi(x_0,\ldots,x_{\ell-1})$.
Then by Lemma~\ref{lem:boff} we have that there are $k\in\mathbb N$ and $(a_{0}^0,\ldots, a_{0}^{\ell-1}),\ldots, (a_{k-1}^0,\ldots, a_{k-1}^{\ell-1})\in \varphi(\A)$
such that 
$$(h(\langle a_{0}^0 a_{1}^0 \cdots a_{k-1}^0\rangle),\ldots,h(\langle
a_{0}^{\ell-1} a_{1}^{\ell-1} \cdots
a_{k-1}^{\ell-1}\rangle))\notin\varphi(\A),$$
 that is,
$$
\big(h(e_k(a_{0}^0, a_{1}^0, \ldots a_{k-1}^0)),\ldots,h(e_k(a_{0}^{\ell-1}, a_{1}^{\ell-1}, \ldots a_{k-1}^{\ell-1}))\big)\notin\varphi(\A).
$$
Hence, $h_{<k}$ is a $k$-ary polymorphism of $\A$ that does not preserve $\varphi$.\proofend

\begin{rem} The converse of (2) is true in case $\A$ satisfies the following condition:
for every surjective periomorphism $h$ of $\A$ there exists $k\in\mathbb N$ such that $h_{<k}$ is surjective. For example, finite structures satisfy this condition.
\end{rem}

We saw that a periomorphism $h$ gives rise to a sequence of polymorphisms $(h_{<k})_{k>0}$. In fact, this gives a one-to-one correspondence with those
polymorphism sequences that satisfy the following property.

\begin{defi} A sequence $(g_k)_{k>0}$ is a {\em cone of polymorphisms of $\A$} if every $g_k$ is a $k$-ary polymorphism of $\A$ and
$g_{\ell}=g_{k}\circ e_{(\ell,k)}$ whenever $\ell< k$ and $\ell$ divides $k$.
\end{defi}

\begin{prop}\label{prop:polycone} A sequence $(g_k)_{k>0}$ is a cone of polymorphisms of $\A$ if and only if there is a periomorphism $h$ of $\A$ such that
$h_{<k}=g_k$ for all $k>0$.
\end{prop}

\proof For the backward direction, let $h$ be a periomorphism of $\A$. Clearly, $(h_{<k})_{k>0}$ is a sequence of polymorphisms of $\A$ -- and it is a cone:
$$
h_{<\ell}=h\circ e_\ell=h\circ (e_{k}\circ e_{(\ell,k)}) =h_{<k}\circ e_{(\ell,k)}.
$$
Here, the second equality follows from the $e_\ell$s being limit embeddings (see the previous section).

Conversely, assume that $(g_k)_{k>0}$ is a cone of polymorphisms of $\A$. Then this is 
a cone of the directed system given by the $e_{(n,m)}$s (viewed as a directed system of homomorphisms). 
By the universal property of limits we get a homomorphism $h$ from $\A^{\textup{per}}\cong \lim_n\mathfrak A^n$ into $\A$ such that $h\circ e_k=g_k$.
\proofend

Intuitively speaking, just as the periodic power is a cone of finite powers, any periomorphism ``is'' a cone of (finitary) polymorphisms.


\section{Preservation theorem}

\begin{thm}[Main]\label{theo:pres} Let $\A$ be an $\aleph_0$-categorical structure. A relation $R$ over $A$ is positive Horn definable in $\A$
if and only if it is preserved by all surjective periomorphisms of~$\A$.
\end{thm}

The following is a straightforward generalization of Proposition~\ref{prop:pres}.

\begin{prop}\label{prop:gpres} If $\A$ and $\B$ are structures such that there is a surjective homomorphism from $\A^{\textup{per}}$ onto $\B$, then $\A\Rrightarrow_{\textup{pH}}\B$.
\qed \end{prop}

The main lemma in the proof of Theorem~\ref{theo:pres} states that a converse of this proposition
holds true in the $\aleph_0$-categorical case:

\begin{lem}\label{lem:main} If $\A$ and $\B$ are $\aleph_0$-categorical structures such that $\A\Rrightarrow_{\textup{pH}}\B$, then there is a surjective homomorphism from $\A^{\textup{per}}$ onto $\B$.
\end{lem}

\noindent{\em Proof:} Let $I$ be the set of finite partial functions $f$ from $\A^{\textup{per}}$ to $\B$ such that
\begin{equation}\label{eq:afbf1}
(\A^{\textup{per}},\bar{\vec a})\Rrightarrow_{\textup{pH}}(\B,\bar b).
\end{equation}
where $\bar{\vec a}$ is a (finite) tuple from $\A^{\textup{per}}$ listing all elements of the domain of $f$ and $\bar b$ is a tuple
from $\B$ such that $f$ maps $\bar{\vec a}$ to $\bar b$.

Observe that $\A^{\textup{per}}$ is countable. Hence, by a standard back and forth argument, it suffices to verify the following two claims.

\medskip

\noindent{\em Claim 1.} For all $f\in I$ and $\vec a\in A^{\textup{per}}$ there is $b\in B$ such that $f\cup\{(\vec a,b)\}\in I$.\medskip

\noindent{\em Claim 2.} For all $f\in I$ and $b\in B$ there is $\vec a\in A^{\textup{per}}$ such that $f\cup\{(\vec a,b)\}\in I$.\medskip

\noindent{\em Proof of Claim 1.} Given $f\in I$ choose a tuples $\bar{\vec a}$ and $\bar b$ as above. Let $\vec a\in A^{\textup{per}}$ be arbitrary. It sufficies to find $b\in B$ such that
\begin{equation}\label{eq:horn}
(A^{\textup{per}},\bar{\vec a},\vec a)\Rrightarrow_{\textup{pH}}(\B,\bar b,b)
\end{equation}
Note in particular that $x=y$ is positive Horn, so \eqref{eq:horn} implies that $f\cup\{(\vec a,b)\}$ is a function. To find such $b$ consider the set
$\Delta(x)$ of all positive Horn formulas $\psi(x)$ (in the language of $(\A^{\textup{per}},\bar{\vec a}$)) satisfied by $\vec a$ in $(\A^{\textup{per}},\bar{\vec a})$. It suffices to 
show this set is satisfiable in $(\B,\bar{b})$.
Since $\B$ is $\aleph_0$-categorical, it is $\aleph_0$-saturated (recall Section~\ref{subsec:cat}), and hence
it suffices to show that every finite subset of $\Delta(x)$ is satisfiable in~$(\B,\bar b)$. 
But for a finite $\Delta_0(x)\subseteq\Delta(x)$ the positive Horn sentence
$\exists x\bigwedge\Delta_0(x)$ is true in $(\A^{\textup{per}},\bar{\vec a})$, so it is also true in $(\B,\bar b)$
by \eqref{eq:afbf1}. Hence $(\B,\bar b)$ contains some $b$ sa\-tis\-fy\-ing~$\Delta_0(x)$.
\hfill$\dashv$\medskip

\noindent{\em Proof of Claim 2.}
Let $f\in I$ and again choose $\bar{\vec a}$ and $\bar b$ as above;
let $k$ denote the length of these tuples.
Again, it suffices given any $b\in B$ to find some $\vec a\in A^{\textup{per}}$ such that \eqref{eq:horn} holds.
As $\A$ is $\aleph_0$-categorical by Ryll-Nardzewski there are up to equivalence in $\A$ only finitely many 
formulas in the variables $\bar yx$ where $\bar y$ is a tuple of $k$ variables. Let
$$
\psi_0(\bar y,x),\ldots,\psi_{m-1}(\bar y,x)
$$
list, up to equivalence in $\A$, all positive Horn formulas $\psi(\bar y,x)$ such that
\begin{equation}\label{eq:bfalse}
\B\not\models\psi(\bar b,b).
\end{equation}


%
%

In particular, for every $j<m$ we have $(\B,\bar b)\not\models\forall x\psi_j(\bar y,x)$ and 
because $f\in I$ also $(\A^{\textup{per}},\bar{\vec a})\not\models\forall x\psi_{j}(\bar y,x)$.
By Lemma~\ref{lem:basic2} there are $i_0\in\mathbb N$ and $a_0\in A$ such that
$$
(\A,\bar{\vec a}(i_0))\not\models\psi_{0}(\bar y,a_0).
$$
Similarly, there are $i_1\in\mathbb N$ and $ a_1\in A$ such that
\begin{equation}\label{eq:i1}
(\A,\bar{\vec a}(i_1))\not\models\psi_{1}(\bar y, a_1).
\end{equation}
Moreover, we can choose $i_1$ such that $i_1>i_0$ by periodicity: if $i_1\le i_0$ replace it by $i_1+i_0\cdot n$ where
$n\in\mathbb N$ is large enough such that all components of $\bar{\vec a}$ are $n$-periodic; then
$\bar{\vec a}(i_1)=\bar{\vec a}(i_1+i_0\cdot n)$ and \eqref{eq:i1} remains true.

Continuing in this manner we get sequences $i_0<i_1<\cdots<i_{m-1}$ and $a_0,a_1,\ldots,a_{m-1}$ such that for all $j<m$
\begin{equation}\label{eq:con}
(\A,\bar{\vec a}(i_j))\not\models\psi_{j}(\bar y,a_j).
\end{equation}
Choose a periodic $\vec a:\mathbb N\to A $ such that for all $j<m$
\begin{equation}\label{eq:defa}
\vec a(i_j)=a_j.
\end{equation}

We verify \eqref{eq:horn} for this $\vec a$: let $\psi(\bar y,x)$ be a positive Horn formula such that $(\B,\bar b)\not\models\psi(\bar y,b)$.
Then there exists $j<m$ such that $\psi(\bar y,x)$ is in $\A$ equivalent to $\psi_j(\bar y,x)$. By \eqref{eq:con} and \eqref{eq:defa} we get
$(\A,\bar{\vec a}(i_j))\not\models\psi_{j}(\bar y,\vec a(i_j))$ and hence $(\A,\bar{\vec a}(i_j))\not\models\psi(\bar y,\vec a(i_j))$. By
Lemma~\ref{lem:basic2} we conclude
$(\A^{\textup{per}},\bar{\vec a})\not\models\psi(\bar y,\vec a)$.\proofend

\noindent{\em Proof of Theorem~\ref{theo:pres}:} The forward direction
follows from Proposition~\ref{prop:pres} (note the $\aleph_0$-categoricity of
$\A$ is not needed).

Conversely, assume that a relation $R\subseteq A^\ell$ is preserved by all surjective perio\-morphisms of $\A$.
By Proposition~\ref{prop:polyper}~(2) it is preserved by all  surjective polymorphisms, and in particular by all automorphisms of $\A$. Since $\A$ is
$\aleph_0$-categorical, $R$ is first-order definable in $\A$ (recall Section~\ref{subsec:cat}). Let $\varphi_R(\bar x)=\varphi_R(x_0,\ldots,x_{\ell-1})$
be a formula such that $R=\varphi_R(\A)$.

By Ryll-Nardzewski there is a finite list of positive Horn formulas
$$
\psi_0(\bar x),\ldots,\psi_{m-1}(\bar x)
$$
in the free variables $\bar x=x_0\cdots x_{\ell-1}$ such that every such formula is in $\A$ equivalent to one from the list. Some of these formulas are
implied  by $\varphi_R(\bar x)$ (in $\A$) and others not, and we may suppose that precisely the first $k$ are not:
\begin{eqnarray}\label{eq:choicetuples}
&&\forall i<k\ \exists \bar a_i\in A^\ell\ : \ \bar  a_i\in\varphi_R(\A)\setminus\psi_i(\A);\\\nonumber
&&\forall k\le j<m\ :\  \varphi_R(\A)\subseteq\psi_j(\A).
\end{eqnarray}
We can assume that $k\neq 0$ as otherwise $(\varphi_R\leftrightarrow\bot)$ holds in $\A$ and then we are done.
We claim that the positive Horn formula $\bigwedge_{k\le j< m}\psi_j(\bar x)$ is equivalent to $\varphi_R(\bar x)$ in $\A$.
Therefore, it suffices to show
$$
\begin{array}{c}
\A\models\forall\bar x\big(\bigwedge_{k\le j< m}\psi_j(\bar x)\to\varphi_R(\bar x)\big).
\end{array}
$$
So we assume that $\bar b$ satisfies  $\bigwedge_{k\le j< m}\psi_j(\bar x)$ in $\A$ and have to show that $\bar b\in\varphi_R(\A)$.

Choose for $i<k$ a tuple $\bar a_i\in A^\ell$ according to~\eqref{eq:choicetuples}.

\medskip

\noindent{\em Claim.} $\prod_{i<k}(\A,\bar a_{i})\Rrightarrow_{\textup{pH}}(\A,\bar b)$.\medskip

\noindent{\em Proof of the claim.} Let $\psi(\bar x)$ be a positive Horn formula that is not satisfied by $\bar b$ in $\A$.
Choose $i<m$ such that $\psi_i(\bar x)$ is equivalent to $\psi(\bar x)$ in $\A$.
Then $\bar b$ does not satisfy $\psi_i(\bar x)$ in $\A$, so $i<k$. But then $(\A,\bar a_{i})\not\models\psi_i(\bar x)$
by \eqref{eq:choicetuples} and thus $(\A,\bar a_{i})\not\models\psi(\bar x)$. As $\psi(\bar x)$ is positive Horn,
$\prod_{i<k}(\A,\bar a_{i})\not\models\psi(\bar x)$ by Lemma~\ref{lem:basic}.\hfill$\dashv$\medskip

Write $\bar a_i=a^0_i\cdots a^{\ell-1}_i$ for $i<k$. Then $\prod_{i<k}(\A,\bar a_{i})$ equals
$$
\big(\A^k, (a^0_{0},\ldots, a^{0}_{k-1})(a^1_{0},\ldots, a^{1}_{k-1})\cdots (a^{\ell-1}_{0},\ldots, a^{\ell-1}_{k-1}) \big).
$$
With $\A$ also $(\A,\bar b)$ is $\aleph_0$-categorical. Further, the structure $\big(\A^k, (a^0_{0},\ldots, a^{0}_{k-1})\cdots\big)$ is $\aleph_0$-catego\-ri\-cal,
because $\A^k$ is (see Section~\ref{subsec:cat}). By the claim we can thus apply Lemma~\ref{lem:main} and
conclude that there is a surjective homomorphism
$$
h:\big(\A^k, (a^0_{0},\ldots, a^{0}_{k-1})\cdots(a^{\ell-1}_{0},\ldots, a^{\ell-1}_{k-1})\big)^{\textup{per}}\twoheadrightarrow (\A,\bar b).
$$
By Proposition~\ref{prop:iso} there is an isomorphism $g$ from the left hand side structure onto
$$
\big(\A^{\textup{per}},\langle a^0_{0}\cdots a^{0}_{k-1}\rangle\cdots\langle a^{\ell-1}_{0}\cdots a^{\ell-1}_{k-1}\rangle\big).
$$
Then $h\circ g^{-1}$ is a surjective homomorphism from $\A^{\textup{per}}$ onto $\A$, i.e.\ a surjective periomorphism of $\A$, such that
$$
h\circ g^{-1}(\langle a^0_{0}\cdots a^{0}_{k-1}\rangle)\cdots h\circ g^{-1}(\langle a^{\ell-1}_{0}\cdots a^{\ell-1}_{k-1}\rangle)=\bar b.
$$
By \eqref{eq:choicetuples} we have $\bar a_i\in\varphi_R(\A)$ for all $i<k$. By Lemma~\ref{lem:boff} and the assumption that $R$ and hence $\varphi_R(\bar x)$
is preserved by surjective periomorphisms of~$\A$, we conclude $\bar b\in\varphi(\A)$, as was to be shown.\proofend


\begin{thm}
\label{cor:surjperio-reduction}
For a finite language $L_0$, let $\B$ be an $L_0$-structure and $\A$ an $L$-structure on the same universe.
If every surjective periomorphism of $\A$ is a periomorphism of $\B$,
then the problem $\qcsp(\B)$ many-one logspace reduces to $\qcsp(\A)$.
\end{thm}

\noindent{\em Proof:} If $\varphi(\bar x)$ is an atomic $L_0$-formula, then $\varphi(\B)$ is preserved by all polymorphisms of $\B$, hence also by
all periomorphisms of $\B$ (by Proposition~\ref{prop:polyper}~(1)), and hence by all surjective periomorphisms of $\A$ (by assumption). By the Main
Theorem~\ref{theo:pres} the relation $\varphi(\B)$ is positive Horn definable in $\A$. Hence $\B$ is positive Horn definable in $\A$. Now apply
Proposition~\ref{prop:qcsp-reduction}.\proofend


\section{Characterization of the pH-hull}

A central tool in constraint complexity is the description of the smallest primitive
positive definable relation containing a given relation $R$ as the smallest relation that contains all polymorphic images of $R$;
this description follows readily from  Theorem~\ref{theo:inv-pol}.
Here we provide a similar tool for quantified constraint complexity.
The proof of this uses most of the results we established so far.

Recall  Definition~\ref{defn:image}.

\begin{thm} Let $\A$ be $\aleph_0$-categorical and
let $R$ be a relation over $A$. Then
\begin{align*}
\{\bar a\mid\ & \exists k\in\mathbb N\ \exists \bar a_0,\ldots \bar a_{k-1}\in R: \bar a  \text{ is a surjective periomorphic image of } (\bar a_i)_{i<k} \}
\end{align*}
is the smallest positive Horn definable relation containing $R$.
\end{thm}

\noindent{\em Proof:} For notational simplicity, we assume that $R$ is binary. It is easy to see that the displayed relation $\tilde R$ contains $R$. We have to show
\begin{enumerate}[(i)]
\item $\tilde R\subseteq\psi(\A)$ for any positive Horn formula $\psi$ such that
$R\subseteq\psi(\A)$;
\item  $\tilde R$ is positive Horn definable in $\A$.
\end{enumerate}
To show (i) let $aa'\in \tilde R$. Choose $a_ia'_i,i<k,$ in $R$
such that some surjective periomorphism of $\A$ maps $\langle a_0\cdots a_{k-1}\rangle\langle a'_0\cdots a'_{k-1}\rangle$ to
$aa'$. Then $a_ia'_i\in\psi(\A)$ as $R\subseteq\psi(\A)$, so $aa'\in\psi(\A)$ by Proposition~\ref{prop:pres} as $\psi$ is positive
Horn.

We now prove (ii). By Theorem~\ref{theo:pres} it suffices to show that $\tilde R$ is
preserved by all surjective periomorphisms of $\A$. We use Lemma~\ref{lem:boff}, so let $a_ia'_i,i<k,$ be $k$ tuples in $\tilde R$ and $h$ be a surjective
periomorphism that maps $\langle a_0\cdots a_{k-1}\rangle\langle a'_0\cdots a'_{k-1}\rangle$ to $aa'$. We have to show that $aa'\in \tilde R$.

For $i<k$ choose $\ell_i$ pairs $b_{ij}b'_{ij},j< \ell_i,$ in $R$ such that there is a surjective periomorphism $h_i$
that maps $\langle b_{i0}\cdots b_{i(\ell_i-1)}\rangle\langle b'_{i0}\cdots b'_{i(\ell_i-1)}\rangle$ to $a_ia'_i$. Letting the
$h_i$s act componentwise we get a surjective homomorphism
\begin{equation}\label{eq:trans}\textstyle
h':\prod_{i<k}(\A^{\textup{per}},\langle b_{i0}\cdots
b_{i(\ell_i-1)}\rangle\langle b'_{i0}\cdots
b'_{i(\ell_i-1)}\rangle)\twoheadrightarrow\prod_{i<k}(\A,a_ia'_i).
\end{equation}
By Proposition~\ref{prop:iso} the left hand side structure is isomorphic to
$$\textstyle
\prod_{i<k}(\A^{\ell_i},(b_{i0}\cdots b_{i(\ell_i-1)})(b'_{i0}\cdots b'_{i(\ell_i-1)}))^{\textup{per}}
$$
and thus by Lemma~\ref{lem:prod} to the periodic power of
$$
\B:=\left(\A^{\sum_{i<k}\ell_i},(b_{00}\cdots b_{(k-1)(\ell_{k-1}-1)}),(b'_{00}\cdots b'_{(k-1)(\ell_{k-1}-1)})\right).
$$
By \eqref{eq:trans} and Proposition~\ref{prop:gpres} we get
\begin{equation}\label{eq:to1}\textstyle
\B\Rrightarrow_{\textup{pH}} \prod_{i<k}(\A,a_ia'_i).
\end{equation}
By Proposition~\ref{prop:iso} the structure
$(\prod_{i<k}(\A,a_ia'_i))^{\textup{per}}$ is isomorphic to the
structure  
$$ (\A^{\textup{per}},\langle a_0\cdots a_{k-1}\rangle,\langle a'_0\cdots a'_{k-1}\rangle) $$
 which maps surjectively onto $(\A,aa')$ by~$h$. Hence, by Proposition~\ref{prop:gpres} again,
\begin{equation}\label{eq:to2}\textstyle
\prod_{i<k}(\A,a_ia'_i)\Rrightarrow_{\textup{pH}} (\A,aa').
\end{equation}
By \eqref{eq:to1} and \eqref{eq:to2} we conclude $\B\Rrightarrow_{\textup{pH}} (\A,aa')$. But these two structures are $\aleph_0$-categorical (by
Ryll-Nardzewski), so Lemma~\ref{lem:main} applies and there is a surjective homomorphism
$$
h'':\B^{\textup{per}}\twoheadrightarrow (\A,aa').
$$
By Proposition~\ref{prop:iso}, $\B^{\textup{per}}$ is isomorphic to
$$
\big(\A^{\textup{per}},\langle b_{00}\cdots b_{(k-1)(\ell_{k-1}-1)}\rangle\langle b'_{00}\cdots b'_{(k-1)(\ell_{k-1}-1)}\rangle\big),
$$
so $aa'$ is a surjective periomorphic image of the $\sum_{i<k}\ell_i$ many pairs
$$
b_{00}b'_{00},\ldots, b_{(k-1)(\ell_{k-1}-1)}b'_{(k-1)(\ell_{k-1}-1)}\in R.
$$
Thus $aa'\in \tilde R$, as was to be shown.\proofend

\section{Equality templates}

Fix a countably infinite set $A$ and define an \emph{equality template} to be a relational structure $\A$ that is
first-order definable in $(A)$, the structure interpreting the empty language; that is, every relation of $\A$ is definable by a pure equality formula.
A complexity classification of the QCSPs of equality templates was given
in previous work~\cite{BodirskyChen10-equality} (see Theorem~\ref{theo:class} below): it was shown that each such QCSP
is either in L, NP-complete or coNP-hard.
In this section, we re-examine this classification theorem.
Based on our Main Theorem~\ref{theo:pres} we give a new proof of this classification which is,
in our view, 
shorter, more modular, and conceptually cleaner than the original proof.

\subsection{Clone analysis} Our proof follows the algebraic approach to constraint complexity and thereby relies on an analysis of
the polymorphism clones of equality templates.  Such clones are locally closed and contain all permutations,
as every permutation of $A$ is an automorphism of $\A$.
Bodirsky, Chen, and Pinsker~\cite{BodirskyChenPinsker10-reducts}, building
on the work of Bodirsky and Kara~\cite{BodirskyKara08-equality},
performed a study of these clones.
Here we state only what we shall need from their analysis.

We define an operation to be \emph{elementary} if it is contained in
the smallest locally closed clone containing all permutations;
a set of operations is \emph{elementary} if each of its operations is elementary.
Let us say that
an operation $f$ {\em generates} another operation $g$ if $g$
is contained in the smallest locally closed clone that contains $f$ and all permutations of $A$.
Note, an operation is elementary if and only
if it is generated by the identity on $A$. Finally, recall that an
{\em essentially unary} operation is one that can be written as the composition of a unary operation and a projection; and, an {\em essential}
operation is one that is not essentially unary.

\begin{lem}[Clone analysis]\label{lem:clone} \
\begin{enumerate}[\em(1)]
\item A non-elementary operation generates either a binary injective operation or a unary constant operation.
\item An operation with infinite image that does not preserve $\neq$ generates all unary operations.
\item Let $k\ge 3$. An essential operation with image size $k$ generates all operations with image size at most $k$.
\end{enumerate}
\end{lem}

\proof The lemma can be derived from results in \cite{BodirskyKara08-equality,BodirskyChenPinsker10-reducts} as follows.
To prove (1), let $f$ be a non-elementary operation.  If $f$ is essentially unary,
then $f$ generates a unary non-elementary operation $h$.
The operation $h$ is not injective, since all unary injective operations
can be interpolated by permutations. By the proof of~\cite[Lemma 10]{BodirskyKara08-equality}, $h$
generates a unary constant operation.

Now suppose that $f$ is essential. By~\cite[Lemma 12]{BodirskyKara08-equality},
$f$ generates an essential binary operation. By~\cite[Theorem 13]{BodirskyKara08-equality},
$f$ generates either  a unary constant operation or a binary injective operation.

Statement (2) follows from \cite[Lemma~38]{BodirskyChenPinsker10-reducts} and statement (3) is
 \cite[Lemma~36]{BodirskyChenPinsker10-reducts}.\proofend

\subsection{Classification} We now start the proof of the classification theorem for equality templates.

\begin{thm}
\label{theo:constant-paste}
Let $\A$ be an equality template such that $\neq$ is not positive Horn definable in $\A$.
Then every unary operation on $A$ is a polymorphism of $\A$.
\end{thm}

\noindent{\em Proof:} If $\neq$ is not positive Horn definable in $\A$, then, by our Main Theorem~\ref{theo:pres}, the relation $\neq$
is not preserved by some surjective periomorphism $h$ of~$\A$.
Recall that according to \eqref{eq:fk} with $h$ there is a naturally associated sequence of polymorphisms $(h_{<k})_{k\ge 1}$.
Because $h$ does not preserve $\neq$, there exists $k_0$
such that $h_{<k_0}$ does not either.
Suppose there exists some $k_1$ such that $h_{<k_1}$ has infinite image. Then $h_{<k_0\cdot k_1}$ does not preserve $\neq$ and has infinite image.
Then our claim follows from Lemma~\ref{lem:clone}~(2). We thus assume that all $h_{<k}$ have finite image.
By local closure it suffices to show:

\medskip

\noindent{\em Claim.} For every $k \in\mathbb N$ every partial unary operation
$g: A \rightarrow A$ that is defined on $k$ points can be extended to a (unary) polymorphism of $\A$.
\medskip

We prove the claim by induction on $k$. For $k = 0$ there is nothing to show.
Suppose that the claim is true for $k$ and let $g$ be a unary operation defined on $k+1$ points.
If $g$ has image size $k+1$, then there exists a permutation
$g'$ extending $g$, and the claim follows;
recall that all permutations are automorphisms of $\A$.
So suppose that $g$ has image of size at most $k$.

It suffices to show that the polymorphism clone of $\A$ contains a unary operation
that has finite image of size $\geq k$, for this implies that the clone contains a unary operation that maps $k+1$ points
to $k$ points; by composing this unary operation with itself and suitable permutations,
one obtains the claim.

Since $h$ has infinite image, there exists $\ell > 0$ such that $h_{<\ell}$ has image size $\geq k$.
Let $\bar a_0, \ldots, \bar a_{k-1} \in A^\ell$ be $k$ many $\ell$-tuples on which $h_{<\ell}$ is injective.
Assume for the sake of notation that $0, \ldots, k-1 \in A$.
Consider the maps $u_0, \ldots, u_{\ell-1}$ defined on $\{ 0, \ldots, k-1 \}$
such that $u_j$ maps each $i<k$ to the $j$th component of $\bar a_i$. Note that $u_0(i)\cdots u_{\ell-1}(i)=\bar a_i$.
By induction every $u_j$ can be extended to a polymorphism $u'_j$ of $\A$.
Define $u: A \rightarrow A$ to map $a\in A$ to $h_{<\ell}(u'_0(a), \ldots, u'_{\ell-1}(a))$. Then
$u(i) = h_{<\ell}(\bar a_i)$ for every  $i <k$, so $u$ is injective on the set $\{ 0, \ldots, k-1 \}$. Thus the image of $u$ has
size $\geq k$ and is finite because it is contained in the image of~$h_{<\ell}$.
\proofend




The following simple lemma will be useful.
It appears as Lemma 11 in~\cite{BodirskyKara08-equality};
we supply a proof for self-containment.

\begin{lem}
\label{lem:constant-or-neq}
Let $\A$ be an equality template.  Either $\A$ has a constant polymorphism,
or the relation $\neq$ is primitive positively definable in $\A$.
\end{lem}

\proof Suppose that $\A$ does not have a constant polymorphism. Then there is a relation $R^{\A}$ that is non-empty and does not contain the constant tuple.
Let $k$ be the arity of $R^{\A}$. Let us say that an equivalence relation $\sigma$ on $\{ 0, \ldots, k-1\}$
is \emph{realized} if there exists a tuple $(a_0 \ldots, a_{k-1})\in R^{\A}$
such that $a_i = a_j$ if and only if $(i, j) \in \sigma$. (Note that if there exists one tuple in $R^{\A}$
satisfying the given condition, then all tuples satisfying the given condition are in $R^{\A}$.)
Let $\tau$ be a coarsest realized equivalence relation.
Consider the relation defined in $\A$ by the primitive positive formula
$$
\begin{array}{c}
\varphi(x_0, \ldots, x_{k-1}) := Rx_0 \cdots x_{k-1} \wedge \bigwedge_{(i, j) \in \tau} x_i = x_j;
\end{array}
$$
in this relation, $\tau$ is realized, and it is the only equivalence
relation that is realized. Since $R^{\A}$ does not contain the constant tuple, $\tau$
contains more than one equivalence class.  Fix $i, j \in \{ 0, \ldots, k-1 \}$ to be values such that $(i, j) \notin \tau$.
The formula $\psi(x_i, x_j)$ derived from $\varphi$ by existentially quantifying
all variables other than $x_i$ and $x_j$ defines the relation $\neq$.
\proofend

Let us say that a relation over $A$ is \emph{negative} if it is definable
as the conjunction of \textit{(i)} equalities and \textit{(ii)} disjunctions of disequalities;
by a disequality, we mean a formula of the form $\neg x = y$.
Let us say that a relation is \emph{positive} if it is definable
using equalities and the binary connectives $\{ \wedge, \vee \}$.
We call an equality template \emph{negative} or \emph{positive} if each of its relations
is negative or positive respectively.

\begin{exa} The ternary relation $P\subseteq A^3$ defined by the formula $\varphi_P(x,y,z):= (x=y\vee y=z)$ in $(A)$ is positive;
it can be verified from the definition that it is not negative.
\end{exa}

\begin{exa} The ternary relation $I\subseteq A^3$ defined by the formula $\varphi_I(x,y,z):= (x=y\to y=z)$ in $(A)$ is neither positive not negative;
this can be verified from the definitions.
\end{exa}

Positivity can be characterized algebraically as follows. This has been shown in~\cite[Proposition 7.3]{BodirskyChen10-equality}.

\begin{prop} \label{prop:char}
Let $\A$ be an equality template,
and fix $f$ to be any non-injective surjective unary operation on $A$.
The following are equivalent:
\begin{enumerate}[--]
\item $\A$ is positive.
\item Every unary operation is a polymorphism of $\A$.
\item The operation $f$ is
a polymorphism of $\A$.
\qed
\end{enumerate}
\end{prop}

We have the following fact.

\begin{cor}
\label{cor:closure-under-phdef}\
\begin{enumerate}[\em(1)]
\item If
$\A$ is a positive equality template, then every positive Horn definable
relation in $\A$ is positive.
\item If $\A$ is a negative equality template, then every positive Horn definable
relation in $\A$ is negative.
\end{enumerate}
\end{cor}

\proof By Proposition~\ref{prop:char} we have that for any fixed
non-injective surjective unary operation $f$, a relation is positive if and
only if it is preserved by $f$; this characterization of positivity
implies (1). 

Likewise, (2) follows from the fact that negativity can be characterized by preservation by a
surjective operation (see \cite[Proposition 68]{BodirskyChenPinsker10-reducts}).\proofend

The following is
known (\cite[Lemma~8.8]{BodirskyChen10-equality}):

\begin{lem}\label{lem:negorI}
 If $R$ is a relation over $A$ that is not negative and is preserved by a binary injective operation, then $I$ is primitive positively definable in $(A,R,\neq)$.
\qed \end{lem}

We are ready to state and prove the classification.

\begin{thm}[\cite{BodirskyChen10-equality}] \label{theo:class}
Let $\A$ be an equality template.
\begin{enumerate}[\em(1)]
\item If $\A$ is negative, then $\qcsp(\A)$ is in $\textup{L}$.
\item If $\A$ is not negative but positive, then the relation $P$ is positive Horn definable in $\A$ and $\qcsp(\A)$ is $\textup{NP}$-complete.
\item If $\A$ is neither negative nor positive, then the relation  $I$ is positive Horn definable in $\A$ and $\qcsp(\A)$ is $\textup{coNP}$-hard.
\end{enumerate}
\end{thm}

\noindent{\em Proof:} We take as given the following complexity results: it is shown in~\cite{BodirskyChen10-equality}  that a negative template $\A$ has $\qcsp(\A)$ in L,
that $\qcsp((A ,P))$ is NP-hard, and that $\qcsp((A, I))$ is coNP-hard; and, it follows from~\cite{Kozen81-positive} that a positive
template $\A$ has $\qcsp(\A)$ in NP.
By Proposition~\ref{prop:qcsp-reduction} and Corollary~\ref{cor:closure-under-phdef},
it thus suffices to show that for an equality template $\A$
one of the following three conditions holds:\begin{enumerate}[(i)]
\item $\A$ is negative.
\item $\A$ is positive and $P$ is positive Horn definable in $\A$.
\item $I$ is positive Horn definable in $\A$.
\end{enumerate}

Let $\A$ be an equality template and let $[\A]_{\textup{pH}}$ denote its expansion by all relations that are positive Horn definable in~$\A$.
Further, let $C$ denote the clone of polymorphisms of $[\A]_{\textup{pH}}$.
By Lemma~\ref{lem:clone}~(1), the following three cases are exhaustive.

\medskip

\noindent{\em Case 1:} $C$ is elementary. Then $C$  preserves $I$, so this relation is primitive positively definable in $[\A]_{\textup{pH}}$ by
Theorem~\ref{theo:inv-pol} and hence positive Horn definable in~$\A$.

\medskip

\noindent{\em Case 2:} $C$ contains a constant operation. Then $\neq$
is not contained in
$[\A]_{\textup{pH}}$, since $\neq$ is not preserved by a constant operation.
Applying Theorem~\ref{theo:constant-paste} to $[\A]_{\textup{pH}}$,
we obtain that $C$ contains all unary operations.
 Proposition~\ref{prop:char} implies
that $[\A]_{\textup{pH}}$ (and hence $\A$) is positive. We claim that either $[\A]_{\textup{pH}}$ (and hence $\A$) is negative or $P$ is positive Horn definable in~$\A$.

{\em Case 2.1:} Suppose that there exists a surjective periomorphism $h$ of $\A$ and a $k>0$ such that
the polymorphism $h_{<k}$ is essential. We claim that in this case $C$ contains all operations.
It is known (and easy to verify) that each relation preserved by this clone can be defined by a conjunction of equalities,
so then $[\A]_{\textup{pH}}$ will be negative.  By local closure,
 it suffices to show that $C$ contains all finite image operations.
Hence, by Lemma~\ref{lem:clone}~(3), it suffices to show that  $C$ contains a sequence of polymorphisms that is \emph{desirable} in the sense that
each polymorphism is essential and has finite image, and that the sequence has unbounded image size.
Now, $(h_{< \ell\cdot k})_{\ell >0}$ is such a desirable sequence in case each $h_{<\ell\cdot k}$ has finite image.
And otherwise there is $\ell_0>0$ such that $h_{< \ell_0\cdot k}$ has infinite image, and then
one obtains a desirable sequence $(u_i\circ h_{< \ell_0\cdot k})_{i>0}$ for suitable unary operations~$u_i$ (recall that all unary operations are in $C$).

{\em Case 2.2:}  Suppose otherwise that for every surjective periomorphism $h$ and all $k>0$ the polymorphism $h_{<k}$ is essentially unary.
We claim that then the relation $P$ is positive Horn definable in $\A$. By our Main Theorem~\ref{theo:pres}
it suffices to show that $P$ is preserved by all surjective periomorphisms of~$\A$. But if a surjective periomorphism $h$ of $\A$ does not preserve $P$,
then there exists $k>0$ such that
$h_{<k}$ does not preserve $P$. Since $h_{<k}$ is essentially unary, this is impossible.

%

\medskip

\noindent{\em Case 3:} $C$ contains a binary injective operation and does not contain a constant operation.
In this case, $[\A]_{\textup{pH}}$ contains $\neq$ by
Lemma~\ref{lem:constant-or-neq}.
It follows immediately from Lemma~\ref{lem:negorI}
that either
$[\A]_{\textup{pH}}$ (and hence $\A$) is negative or
 $I$ is primitive positively definable in $[\A]_{\textup{pH}}$ and hence
positive Horn definable in~$\A$.
\proofend

\section{Discussion}

Bing's theorem~\cite{bing} involves a clever, technical argument
that allows us to streng\-then our main preservation theorem for
structures that are isomorphic to their finite powers.
Such structures have gained some attention in constraint complexity \cite{BCKV09-maximal,BodirskyHilsMartin10-scopeunivalg}. We have the following theorem.

\begin{thm}\label{theo:ae} Let $\A$ be a countable $\aleph_0$-categorical structure such that $\A\cong\A^2$.  Then a formula $\varphi(\bar x)$ is equivalent to a positive Horn formula in $\A$ if and only if it is preserved by all surjective polymorphisms of $\A$.
\end{thm}

\proof
Let $\A$ accord the assumption of the theorem.
We only prove the backward direction.
Assume $\varphi(\bar x)$ is preserved by all surjective polymorphisms of $\A$.
In particular, $\varphi(\bar x)$ is preserved by all surjective homorphisms from $\A$ to $\A$.
It is not hard to see that Lyndon's Theorem implies that there exists a  positive formula $\varphi^+(\bar x)$ such that
 $\varphi(\A)=\varphi^+(\A)$  (see \cite[Proposition 2 (c)]{junker} for details). We can assume that $\varphi^+$ has
the form of some quantifier prefix followed by a quantifier free formula
$$\textstyle
\psi=\bigwedge_{i\in I}\bigvee_{j\in J}\alpha_{ij},
$$
where the $\alpha_{ij}$s are atoms. For each $f\in J^{I}$ write
$$\textstyle
\psi_f:=\bigwedge_{i\in I}\alpha_{if(i)}.
$$

\noindent{\em Bing's argument.} Let $\bar Q\bar y$ be an arbitrary quantifier prefix. Assume for every $f\in J^{I}$ the tuple $\bar a_f$ in $\A$ is an assignment to the free variables in $\bar Q\bar y\psi$ such that $\prod_{f\in J^{I}}(\A,\bar a_f)\models\bar Q\bar y\psi$. Then there exists $f\in J^{I}$ such that $(\A,\bar a_f)\models\bar Q\bar y \psi_f$.\medskip

\noindent{\em Proof of Bing's argument.} This can be proved by a straightforward induction on the length of $\bar Q\bar y$. See \cite[Lemma~3]{bing} for details.\hfill$\dashv$\medskip

Write $\varphi^+(\bar x)=\bar Q\bar y\psi(\bar y,\bar x)$. \medskip

\noindent{\em Claim.} There exists $f\in J^{I}$ such that $\A\models\forall\bar x(\varphi^+(\bar x)\to\bar Q\bar y\psi_f(\bar y, \bar x))$.\medskip

\noindent{\em Proof of Claim.} Otherwise we find for every $f\in J^{I}$ an $\bar a_f\in\varphi^+(\A)$ such that
$$
(\A,\bar a_f)\not\models \bar Q\bar y\psi_f(\bar y, \bar x).
$$
Then $\prod_{f\in J^{I}}(\A,\bar a_f)\not\models\varphi^+(\bar x)$ by Bing's argument.
As
$\A\cong\A^2$, there is an isomorphism
$$
h: \A^{J^{I}}\cong \A.
$$
Write $\bar x=x_0\cdots x_{\ell-1}$ and $\bar a_f=a_f^{0}\cdots a_{f}^{\ell-1}$.
Then
\begin{align*}\textstyle
h:\prod_{f\in J^{I}}(\A,\bar a_f)&=\big(\A^{J^{I}},(a_f^{0})_{f\in J^{I}},\ldots,(a_f^{\ell-1})_{f\in J^{I}} \big)\\
&\cong\big(\A, h((a_f^{0})_{f\in J^{I}}),\ldots,h((a_f^{\ell-1})_{f\in J^{I}})\big).
\end{align*}
Since $h$ is an isomorphism, $\varphi^+(\bar x)$ is false in the right hand side structure.
Hence $h$ is (up to a renaming of indices) a surjective polymorphism of $\A$ that does not preserve $\varphi(\bar x)$, a contradiction.\hfill$\dashv$\medskip

Since $(\bar Q\bar y\psi_f\to\varphi^+)$ is logically valid, the claim implies that $\varphi^+$ is equivalent in $\A$ to the positive Horn formula $\bar Q\bar y\psi_f$.\proofend

\begin{exas} An example of a structure satisfying the assumption of the theorem is the countable atomless Boolean algebra (cf.~\cite[Section~5.2]{bodsurvey}). This template is of central importance for spatial reasoning in artificial intelligence. Another example is an infinite dimensional vectorspace over some finite field (cf.~\cite[Example~2.10]{BBCJK09-qcsp},~\cite[Section~5.3]{bodsurvey}). More generally, it is easy to see that every countable $\aleph_0$-categorical structure $\A$ whose theory is Horn axiomatizable satisfies $\A\cong\A^2$.
\end{exas}

We conclude with some remarks and questions.

\medskip

\noindent $\bullet$
Very recently,
Bodirsky, Hils and Martin~\cite{BodirskyHilsMartin10-scopeunivalg} explored the possibilities to
extend the algebraic machinery for constraint satisfaction
 to structures that are not necessarily
$\aleph_0$-categorical; they established a variant of the
preservation theorem for primitive positive definability via
$\omega$-polymorphisms for structures that are in a certain sense
sufficiently saturated. (An $\omega$-polymorphism of a structure $\A$
is a homomorphism from $\A^{\mathbb N}$ to~$\A$.)

\medskip

\noindent $\bullet$
The first author showed \cite[Lemma~7.5]{Chen09-Rendezvous} that, in finite structures, positive Horn definability coincides with $\Pi_2$ positive Horn definability (see \cite{mm,ChenMadelaineMartin08-containment} for a related result). Using the method of the proof,
one can infer that Boolean QCSPs with quantifier alternation rank
restricted to some
even $t\ge 2$ are either $\Pi_t^{\textup{P}}$-complete or in P (cf.~\cite[Theorem~7.2]{Chen09-Rendezvous}).
An open issue is to study $\aleph_0$-categorical QCSPs with
bounded alternation rank.

One can ask the following concrete question.
Let $\A$ be a $\aleph_0$-categorical structure and $\varphi$ a $\Pi_t$ formula that is preserved by the surjective periomorphisms of $\A$. Is $\varphi$ equivalent to a positive Horn formula that is also $\Pi_t$?

\medskip

\noindent $\bullet$
A related question is posed by Y. Chen and Flum in \cite{flumchen}. They ask for an alternation rank preserving version of Lyndon's preservation theorem: is any $\Pi_t$ sentence that is preserved by surjective homomorphisms equivalent to a positive $\Pi_t$ sentence? This is known to be true for $t\le 2$ \cite{ritter}.
By a well-known trick of Lyndon \cite{lyndon} (see also Fefermann's survey~\cite{fefer}) a positive answer would follow from
a proof of the following:
any implication between $\Pi_t$ formulas has a $\Pi_t$ Lyndon-interpolant.
 The usual argument constructs an interpolant by recursion on a cut-free proof of the given implication. But again for $t>3$ there seems to be no control on the alternation rank of an interpolant constructed in this way.

\section*{Acknowledgments} Manuel Bodirsky and Barnaby Martin made valuable comments on an early version of this paper.
The authors also thank Manuel for useful literature pointers.



%




\bibliographystyle{plain}


\bibliography{moritzbib,hubiebib}

\begin{thebibliography}{10}

\bibitem{ABISV09-refining}
E.~Allender, M.~Bauland, N.~Immerman, H.~Schnoor, and H.~Vollmer.
\newblock {The Complexity of Satisfiability Problems: Refining Schaefer's
  Theorem.}
\newblock {\em Journal of Computer and System Sciences}, 75(4):245--254, 2009.

\bibitem{BartoKozik09-boundedwidth}
L.~Barto and M.~Kozik.
\newblock Constraint satisfaction problems of bounded width.
\newblock In {\em Proceedings of FOCS'09}, 2009.

\bibitem{bing}
K.~Bing.
\newblock On arithmetical classes not closed under direct union. Proceedings of
  the American Mathematical Society 6:836-846, 1955.

\bibitem{bodsurvey}
M.~Bodirsky.
\newblock Constraint satisfaction problems with infinite templates. In N.
  Creignou et al. (eds.), Complexity of Constraints - An Overview of Current
  Research Themes, LNCS 5250, pp. 196-228, 2008.

\bibitem{BodirskyHermannRichoux09-existentialpositive}
M.~Bodirsky, M.~Hermann, and F.~Richoux.
\newblock Complexity of existential positive first-order logic.
\newblock In {\em Proceedings of Computability in Europe}, pages 31--36, 2009.

\bibitem{BodirskyHilsMartin10-scopeunivalg}
M.~Bodirsky, M.~Hils, and B.~Martin.
\newblock On the scope of the universal-algebraic approach to constraint
  satisfaction.
\newblock In {\em Proceedings of the 25th IEEE Symposium on Logic in Computer
  Science}, 2010.

\bibitem{junker}
M.~Bodirsky and M.~Junker.
\newblock Aleph0-categorical structures: interpretations and endomorphisms.
  Algebra Universalis 64(3-4):403-417, 2010.

\bibitem{Bodirsky04-thesis}
Manuel Bodirsky.
\newblock {Constraint Satisfaction with Infinite Domains}.
\newblock PhD thesis, Humboldt-Universitat zu Berlin, 2004, 2004.

\bibitem{BodirskyChen10-equality}
Manuel Bodirsky and Hubie Chen.
\newblock Quantified equality constraints.
\newblock {\em SIAM Journal on Computing}, 39(8):3682--3699, 2010.

\bibitem{BCKV09-maximal}
Manuel Bodirsky, Hubie Chen, Jan Kara, and Timo von Oertzen.
\newblock Maximal infinite-valued constraint languages.
\newblock {\em Theoretical Computer Science}, 410:1684--1693, 2009.

\bibitem{BodirskyChenPinsker10-reducts}
Manuel Bodirsky, Hubie Chen, and Michael Pinsker.
\newblock The reducts of equality up to primitive positive interdefinability.
\newblock {\em Journal of Symbolic Logic}, 75(4):1249--1292, 2010.

\bibitem{BodirskyKara08-equality}
Manuel Bodirsky and Jan K\'ara.
\newblock The complexity of equality constraint languages.
\newblock {\em Theory of Computing Systems}, 3(2):136--158, 2008.
\newblock A conference version appeared in the proceedings of CSR'06.

\bibitem{BodirskyNesetril06-homogeneous}
Manuel Bodirsky and Jaroslav Ne\v{s}et\v{r}il.
\newblock Constraint satisfaction with countable homogeneous templates.
\newblock {\em Journal of Logic and Computation}, 16(3):359--373, 2006.

\bibitem{BKKR69-galois}
V.~G. Bodnar\v{c}uk, L.~A. Kalu\v{z}nin, V.~N. Kotov, and B.~A. Romov.
\newblock Galois theory for post algebras, part {I} and {II}.
\newblock {\em Cybernetics}, 5:243--252, 531--539, 1969.

\bibitem{BBCJK09-qcsp}
Ferdinand B\"{o}rner, Andrei~A. Bulatov, Hubie Chen, Peter Jeavons, and
  Andrei~A. Krokhin.
\newblock The complexity of constraint satisfaction games and {QCSP}.
\newblock {\em Information and Computation}, 207(9):923--944, 2009.

\bibitem{ChandraMerlin77-optimal}
Ashok~K. Chandra and Philip~M. Merlin.
\newblock Optimal implementation of conjunctive queries in relational data
  bases.
\newblock In {\em Proceddings of {STOC}'77}, pages 77--90, 1977.

\bibitem{keisler}
C.~C. Chang and H.~J. Keisler.
\newblock Model Theory. Studies in Logic and the Foundations of Mathematics 73.
  North-Holland Publishing Co., Amsterdam, third edition, 1990.

\bibitem{Chen08-collapsibility}
Hubie Chen.
\newblock {The Complexity of Quantified Constraint Satisfaction:
  Collapsibility, Sink Algebras, and the Three-Element Case}.
\newblock {\em {SIAM} Journal on Computing}, 37(5):1674--1701, 2008.

\bibitem{Chen09-Rendezvous}
Hubie Chen.
\newblock A rendezvous of logic, complexity, and algebra.
\newblock {\em {ACM} Computing Surveys}, 42(1), 2009.

\bibitem{Chen11-qcspandpgp}
Hubie Chen.
\newblock Quantified constraint satisfaction and the polynomially generated
  powers property.
\newblock {\em Algebra Universalis}, 65:213--241, 2011.

\bibitem{Chen12-meditations}
Hubie Chen.
\newblock Meditations on quantified constraint satisfaction.
\newblock In Robert Constable and Alexandra Silva, editors, {\em Logic and
  Program Semantics}, volume 7230 of {\em Lecture Notes in Computer Science},
  pages 35--49. Springer Berlin / Heidelberg, 2012.

\bibitem{ChenMadelaineMartin08-containment}
Hubie Chen, Florent Madelaine, and Barnaby Martin.
\newblock Quantified constraints and containment problems.
\newblock In {\em Twenty-Third Annual IEEE Symposium on Logic in Computer
  Science (LICS)}, 2008.

\bibitem{ChenMueller12-preservation}
Hubie Chen and Moritz M\"{u}ller.
\newblock An algebraic preservation theorem for aleph-zero categorical
  quantified constraint satisfaction.
\newblock In {\em ACM/IEEE Symposium on Logic in Computer Science}, 2012.

\bibitem{flumchen}
Y.~Chen and J.~Flum.
\newblock The parameterized complexity of maximality and minimality problems.
  Proceedings of the 2nd International Workshop on Parameterized and Exact
  Computation, pp. 25-37, 2006.

\bibitem{FederVardi99-structure}
T.~Feder and M.~Vardi.
\newblock The computational structure of monotone monadic {SNP} and constraint
  satisfaction: {A} study through {D}atalog and group theory.
\newblock {\em {SIAM} Journal on Computing}, 28:57--104, 1999.

\bibitem{fefer}
S.~Fefermann.
\newblock Harmonious logic: Craig's interpolation theorem and its descendants.
  Synthese 164:341-357, 2008.

\bibitem{flum}
J.~Flum.
\newblock First order logic and its extensions. In G.H. M\"{u}ller et al.
  (eds.), Logic Conference Kiel 1974, Lecture Notes in Mathematics 499, 1975.

\bibitem{Geiger68-closed}
D.~Geiger.
\newblock {Closed Systems of Functions and Predicates}.
\newblock {\em Pacific Journal of Mathematics}, 27:95--100, 1968.

\bibitem{IMMVW10-tractabilityfewsubpowers}
P.~Idziak, P.~Markovic, R.~McKenzie, M.~Valeriote, and R.~Willard.
\newblock Tractability and learnability arising from algebras with few
  subpowers.
\newblock {\em {SIAM} J. Comput.}, 39(7):3023--3037, 2010.

\bibitem{a}
H.~J. Keisler.
\newblock Reduced products and Horn classes. Transactions of the American
  Mathematical Society 117:307-328, 1965.

\bibitem{Kozen81-positive}
Dexter Kozen.
\newblock Positive first-order logic is {NP}-complete.
\newblock {\em {IBM} Journal of Research and Development}, 25(4):327--332,
  1981.

\bibitem{Krasner68-endomorphisms}
M.~Krasner.
\newblock Endoth\'eorie de {G}alois abstraite.
\newblock {\em S\'eminaire P. Dubreil (Alg\'ebre et Th\'eorie des Nombres)},
  1(6), 1968.

\bibitem{LaroseTesson09-hardness}
Benoit Larose and Pascal Tesson.
\newblock Universal algebra and hardness results for constraint satisfaction
  problems.
\newblock {\em Theoretical Computer Science}, 410(18):1629--1647, 2009.

\bibitem{lyndon}
R.C. Lyndon.
\newblock Properties preserved under homomorphism. Pacific Journal of
  Mathematics 9(1):143-154, 1959.

\bibitem{mm}
F.~Madelaine and B.~Martin.
\newblock The preservation properties of positive Horn logic. Manuscript,
  available at \texttt{www.dur.ac.uk/barnaby.martin/publications.html}, 2009.

\bibitem{MadelaineMartin09-complexityPFOwithouteq}
Florent Madelaine and Barnaby Martin.
\newblock The complexity of positive first-order logic without equality.
\newblock In {\em 24th Annual IEEE Symposium on Logic In Computer Science},
  pages 429--438, 2009.

\bibitem{MadelaineMartin11-tetrachotomyPFOwithouteq}
Florent Madelaine and Barnaby Martin.
\newblock A tetrachotomy for positive first-order logic without equality.
\newblock In {\em 26th Annual IEEE Symposium on Logic In Computer Science},
  pages 311--320, 2011.

\bibitem{Martin08-FOparameterizedbymodel}
Barnaby Martin.
\newblock First-order model checking problems parameterized by the model.
\newblock In {\em Conference on Computability in Europe ({CiE})}, pages
  417--427, 2008.

\bibitem{Papadimitriou95-complexity}
C.H. Papadimitriou.
\newblock {\em {Computational Complexity}}.
\newblock Addison-Wesley, 1995.

\bibitem{ritter}
C.~Ritter.
\newblock Fagin-Definierbarkeit. Diplomarbeit, Universit\"at Freiburg, 2005.

\bibitem{vogler}
H.~Vogler.
\newblock A unifying approach to theorems on preservation and interpolation for
  binary relations between structures. Archive of Mathematical Logic 21(1):
  101-112, 1981.

\end{thebibliography}


\end{document}